\documentclass[review,12pt]{elsarticle}

\usepackage{tabularx} % Ensure this is included in the preamble
\usepackage{lineno} % The 'switch' option places line numbers on the outer margin.
\usepackage{epsfig}
\usepackage{caption}
\usepackage{epstopdf}
\usepackage{graphicx}
\usepackage{amssymb} % The amssymb package provides various useful mathematical symbols
\usepackage{amsthm} % The amsthm package provides extended theorem environments
\usepackage{amsmath}
\usepackage{mathrsfs}
\usepackage{graphicx}
\usepackage{epstopdf}
\usepackage{float}
\usepackage{caption}
\usepackage{soul}
\usepackage{color,soul} %to highlight text
\usepackage{color} %To highlight references in the text
\usepackage{subcaption}
\usepackage{bm}
\usepackage{bbm}
\usepackage{mathrsfs}
\usepackage{soul}
\usepackage[margin=2cm]{geometry}% to reduce margins
\usepackage{booktabs}
\usepackage{chngpage}
\usepackage{enumitem}%enumeration
\usepackage{subcaption}
\usepackage{multirow}
\usepackage{stackengine} %overlap plots
\usepackage[version=4]{mhchem} %chemical formulas
\usepackage{hyperref} %links
\usepackage{cleveref}
\usepackage{textcomp} 
\usepackage{siunitx}
\usepackage[utf8]{inputenc}
\usepackage[T1]{fontenc}

\biboptions{square,comma,sort&compress}
\usepackage{amssymb}
\usepackage{booktabs}
\usepackage{textcomp} 
\usepackage{siunitx}
\usepackage{graphicx}
\usepackage{subcaption}
\usepackage{amsmath}
\usepackage{array}
\usepackage{gensymb}
\usepackage{float}
\usepackage[utf8]{inputenc}
\usepackage{tikz}
\usepackage{float}
\usepackage[numbers,sort&compress]{natbib}
\journal{ }

\makeatletter
\def\@author#1{\g@addto@macro\elsauthors{\normalsize%
    \def\baselinestretch{1}%
    \upshape\authorsep#1\unskip\textsuperscript{%
      \ifx\@fnmark\@empty\else\unskip\sep\@fnmark\let\sep=,\fi
      \ifx\@corref\@empty\else\unskip\sep\@corref\let\sep=,\fi
      }%
    \def\authorsep{\unskip,\space}%
    \global\let\@fnmark\@empty
    \global\let\@corref\@empty  %% Added
    \global\let\sep\@empty}%
    \@eadauthor={#1}
}
\makeatother

\makeatletter
\def\thickhline{%
  \noalign{\ifnum0=`}\fi\hrule \@height \thickarrayrulewidth \futurelet
   \reserved@a\@xthickhline}
\def\@xthickhline{\ifx\reserved@a\thickhline
               \vskip\doublerulesep
               \vskip-\thickarrayrulewidth
             \fi
      \ifnum0=`{\fi}}
\makeatother

\newlength{\thickarrayrulewidth}
\begin{document}

\begin{frontmatter}



\title{New insights into hydrogen-assisted intergranular cracking in nickel}


\author{Sangchu Quan\fnref{IC,OX}}

\author{Alfredo Zafra\fnref{OX}}

\author{Emilio Mart\'{\i}nez-Pa\~neda\fnref{OX}}

\author{Chao Wu\fnref{IC}}

\author{Zachary D. Harris\fnref{Pitt}}

\author{Livia Cupertino-Malheiros\corref{cor1}\fnref{IC}}
\ead{l.cupertino-malheiros@imperial.ac.uk}

\address[IC]{Department of Civil and Environmental Engineering, Imperial College London, London SW7 2AZ, UK}

\address[OX]{Department of Engineering Science, University of Oxford, Oxford OX1 3PJ, UK}

\address[Pitt]{Department of Mechanical Engineering and Materials Science, University of Pittsburgh, Pittsburgh, PA 15261, USA}

\cortext[cor1]{Corresponding author.}

\begin{abstract}

We characterize the grain boundary (GB) susceptibility to hydrogen-assisted intergranular cracking in pure nickel as a function of coincident site lattice value ($\Sigma$-n), over a wide range of hydrogen concentrations (4 to 14 wppm). Cracks on the surface and within the bulk material were identified across the entire gauge region of the specimens. The susceptibility of GBs to crack initiation and propagation was evaluated by separating cracks containing single GB or multiple GBs. A larger loss in fracture strain, a smaller reduction in area, and an increase in the percentage of intergranular fracture indicated a higher degree of embrittlement at elevated hydrogen concentrations. The number of cracks was significantly higher on the surface than in the bulk for the most severe hydrogen charging conditions ($\geq$ 8 wppm), while a similar number was observed for lower concentrations. The propensity for hydrogen-assisted intergranular cracking at different types of GBs on the surface and in the bulk material was consistent, indicating that while cathodic charging can promote surface cracks, it does not significantly impact the GBs relative susceptibility. The $\Sigma$-3 boundaries were the most resistant to cracking, as evidenced by the considerably lower fraction of these GBs exhibiting intergranular cracking at all hydrogen concentrations considered. This contrasts literature findings for Ni alloys and can be explained by the segregation energies and reductions in the cohesive strength with hydrogen, with less favorable trapping at the $\Sigma$-3 boundaries. No evidence of plasticity-mediated cracking initiation was observed.\\
\end{abstract}

\begin{keyword}

Hydrogen embrittlement \sep Intergranular cracking \sep Grain boundary engineering  \sep Nickel



\end{keyword}
\end{frontmatter}

%\linenumbers
\section{Introduction}
\label{sec:Introduction}

In an expanding hydrogen-based economy, it is expected that many metallic structures and components will be exposed to hydrogen-containing environments. This poses a technological challenge. Due to its small atomic size, hydrogen can easily be absorbed into metals, degrading their mechanical properties and ultimately resulting in premature brittle failure through a phenomenon known as hydrogen embrittlement (HE) \cite{Li2020, Donovan1976, Martin2012,Ferreira1998}. Grain boundaries are a common fracture path in this brittle failure mode, with hydrogen-assisted intergranular cracking reported in many metals and alloys under various hydrogen charging conditions \cite{Lu2020, Kumar2017, Tabata1984, Djukic2019, Martin2012, Singh2024MatSciA, Soundararajan2023, Liu2024MatSciA}. Therefore, there has been considerable interest in understanding the HE susceptibility of GBs \cite{Robertson1984, Martin2012, Harris2018, Bechtle2009}. GBs can be preferential hydrogen trapping sites, depending on their character \cite{Lee1986, Chen2024, Oudriss2012}. For example, this is the case for GBs with a binding energy of 20-30 \SI{}{\kilo\joule\per\mole} in pure nickel \cite{Wada2023, Wada2019}. In addition, hydrogen transported by dislocation movement and pile-ups might also contribute to hydrogen accumulation at GBs \cite{Martin2012, Hachet2021}. The accumulation of hydrogen at these interfaces reduces their cohesive energy, weakening their strength, and favoring fracture. For example, Harris \textit{et al.} \cite{Harris2018} observed hydrogen-induced intergranular microcracks in nickel tensile tested at 77 \SI{}{\kelvin}, showing that even at temperatures where hydrogen-plasticity interactions are suppressed, hydrogen accumulation at GBs can lead to cracking.

Identifying the resistance of the type(s) of GBs to hydrogen-induced intergranular cracking is crucial to guide alloy design \cite{Jothi2016, Sangid2010, Harris2019MatSciA}. GBs can be characterized by the number of atoms included in their coincident site lattice (CSL), referred to as the $\Sigma$ value according to Brandon's criterion \cite{Brandon1966}. This geometrical atomic arrangement, characterized by $\Sigma$, results in different mechanical and thermodynamic properties, such as interface energy and hydrogen solubility. Fracture can occur once the hydrogen coverage is sufficient to induce a certain reduction in the GB interface energy. Therefore, it is important to investigate the susceptibility of GBs to HE based on their $\Sigma$ value. Attempts have been made to identify which GBs are more resistant to hydrogen. Studies using density functional theory and molecular dynamics have shown that GBs have different interface energies, both without and with hydrogen, depending on their $\Sigma$ value \cite{Tehranchi2017-II, Mai2021, Li2020DFT, DiStefanoDavide2015, Alvaro2015, Hajilou2020, Li2019, Ding2022}. The strength of GBs with different orientations was also evaluated using micro-bending tests \cite{Alvaro2015, Takahashi2016}. Despite these attempts to obtain the fracture strength of grain boundaries with hydrogen, the results are still scarce, with a large scatter between the different studies \cite{Tehranchi2017-II, Mai2021, Li2020DFT, DiStefanoDavide2015, Alvaro2015, Li2019, Hajilou2020}. For example, the $\Sigma$-3 and $\Sigma$-5 GBs are commonly studied, while other boundaries (e.g., $\Sigma$-29) have rarely been investigated, and their susceptibility to hydrogen-assisted intergranular cracking remains unclear. 

A statistical approach was previously used to evaluate the resistance of GBs to hydrogen, which suggested that $\Sigma$-3 GBs are more prone to hydrogen-assisted cracking in Inconel 725 \cite{Seita2015}. Additional studies \cite{Zhang2020, Liu2024} have also reported increased susceptibility for $\Sigma$-3 GBs and their adjacent regions to HE in Ni alloys. Despite extensive discussions on hydrogen-dislocation interactions in these works, the underlying mechanism responsible for hydrogen promoting $\Sigma$-3 GB fracture remains unresolved. Seita \textit{et al.} \cite{Seita2015} attributed the fracture to suppressed dislocation transmission, resulting in the formation of dislocation substructures within the $\Sigma$-3 GB plane. Liu \textit{et al.} \cite{Liu2024} observed that cracks were more likely to initiate near slipping $\Sigma$-3 GBs, but without concluding that slip is the dominant reason for their weakness. Zhang \textit{et al.} \cite{Zhang2020} identified hydrogen-assisted cracks and dislocation slip bands forming at a `weak' trilayer interface due to $\gamma''$ approximately 40 \SI{}{\nano\meter} away from the $\Sigma$-3 GB, rather than directly on the GB. Therefore, the inherent structural complexity of nickel alloys has posed challenges to fully understanding the resistance of GBs to hydrogen. The presence of second-phase particles, such as $\gamma'$ and $\gamma''$, obfuscates what is happening at the grain boundaries themselves. There is thus a strong need for studying the susceptibility of grain boundaries to hydrogen cracking in a pure material where no such complications exist.  

This study provides a comprehensive analysis of the susceptibility of GBs to hydrogen-assisted cracking in pure nickel at various hydrogen concentrations. The fractured GBs were systematically categorized to elucidate their relative susceptibility to hydrogen-induced cracking. The results consistently reveal that $\Sigma$-3 occurrence among the fractured GBs is significantly lower than in the baseline microstructure, while other $\Sigma$-low GBs accounted for an intermediate fraction of fractures. The orientations of the fractured GBs are predominantly perpendicular to the loading axis, revealing the important contribution of normal stress to hydrogen-assisted intergranular cracking. In contrast to studies on nickel alloys \cite{Zhang2020, Seita2015, Liu2024}, this work purposely removed the influence of second-phase particles to study the relative susceptibility of different grain boundary types. The results emphasize the remarkable resistance of $\Sigma$-3 GBs to hydrogen-assisted intergranular cracking under fracture conditions predominantly driven by decohesion.

\section{Methodology}
\label{sec:Methodology}

\subsection{Material and baseline characterization}

\begin{figure}[th]
    \centering
    \includegraphics[width=1.0\textwidth]{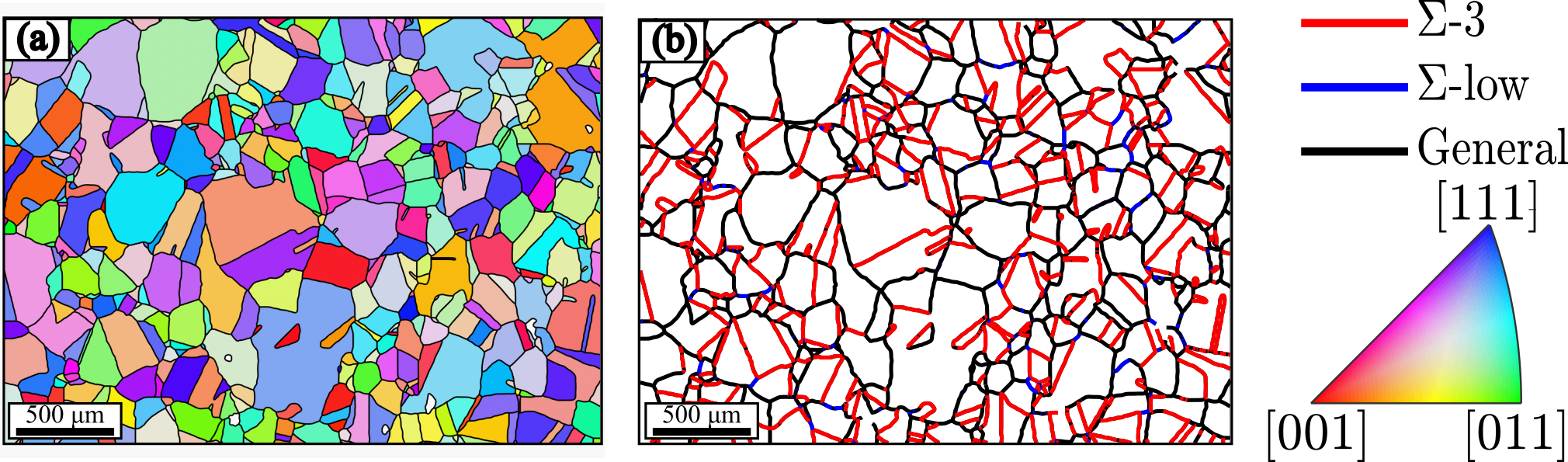}  % Adjust width as needed
    \caption{Baseline microstructure: (a) IPF map and (b) three GB categories showing $\Sigma$-3 in red, other $\Sigma$-low in blue and general in black.}
    \label{fig:Baseline microstructure.}
\end{figure}

All experiments were performed on 99.9 wt\% pure nickel procured as a 20 \SI{}{\milli\meter} thick cold-rolled plate. The as-received material was annealed at 950 \SI{}{\celsius} for 1 hour, resulting in the baseline microstructure shown in Fig. \ref{fig:Baseline microstructure.}, with equiaxed grains and an average size of 101 $\pm$ 7 \SI{}{\micro\meter}.  

Samples of 10 $\times$ 10 $\times$ 0.7 \SI{}{\cubic\milli\meter} and of 8 \SI{}{\milli\meter} gauge length with 3 $\times$ 0.7 \SI{}{\square\milli\meter} cross-section were electrical discharge machined for thermal desorption spectroscopy (TDS) analysis and tensile tests, respectively. The surfaces of both samples were ground up to 2000 grit (7 \SI{}{\micro\meter}) with SiC paper. The tensile specimens were then polished with 6, 3, 1, and 0.25 \SI{}{\micro\meter} diamond suspensions and finished with a 0.25 \SI{}{\micro\meter} OPS silica suspension to achieve the high-quality surface necessary for electron backscatter diffraction (EBSD). 

The $\Sigma$ value of GBs in the gauge region was characterized following the criteria described in \ref{appendix}, as shown in Fig. \ref{fig:Baseline microstructure.}. The GBs were then classified into $\Sigma$-3, $\Sigma$-low (3 < $\Sigma \leq 29$) and general GBs ($\Sigma > 29$, and random GBs). The fractions of these three groups, given in terms of the number of GB segments (where each GB segment has identical misorientation), reflect the population of GBs in the annealed specimens. These fractions, obtained after heat treatment but before deformation, are considered the baseline microstructure.

\subsection{Electrochemical charging and TDS analyses}

After this baseline characterization, the specimens were then electrochemically charged using a Gamry 1010B potentiostat, with Ag/AgCl as the reference electrode and Pt as the counter electrode. To achieve a wide range of hydrogen concentrations, 0.6 \SI{}{\mole\per\liter} NaCl and 0.1 \SI{}{\mole\per\liter} NaOH solutions were employed with applied constant cathodic current densities of -4 \SI{}{\milli\ampere\per\centi\meter\squared} and -15 \SI{}{\milli\ampere\per\centi\meter\squared}. Electrochemical charging was performed for 70 hours at 70 \SI{}{\celsius}. This condition was chosen to reach a uniform hydrogen concentration throughout the sample with a concentration of 98\% at the middle thickness, considering a 1D diffusion model and a hydrogen diffusivity in Ni of $8.32 \times 10^{-13}$ \SI{}{\square\meter\per\second} \cite{Lee1986}. 

TDS measurements were conducted with a Hiden Analytical RC PIC quadrupole mass spectrometer to determine the hydrogen concentration after electrochemical charging. Two samples were tested for each charging condition. The measurements were performed with a heating rate of 30 \SI{}{\celsius\per\minute} and a current of 20 \SI{}{\micro\ampere} up to 1000 \SI{}{\celsius}. 

\subsection{Tensile testing and hydrogen-assisted intergranular cracking}

Uniaxial tensile tests at a strain rate of $2 \times 10^{-4}$ \SI{}{\second^{-1}} were performed with a Deben MT1000 Microtest stage. Two specimens were tested for each hydrogen charging condition, including an uncharged specimen. To ensure comparable hydrogen concentrations, both the TDS and the tensile tests were started 30 minutes after the end of the hydrogen charging. The fracture strain obtained from the stress-strain response curve was used as an indicator of HE. In addition, the total fracture surface area and the fraction of the intergranular fracture area were also measured to provide a further assessment of HE levels.

\begin{figure}[th]
    \centering
    \includegraphics[width=1.0\textwidth]{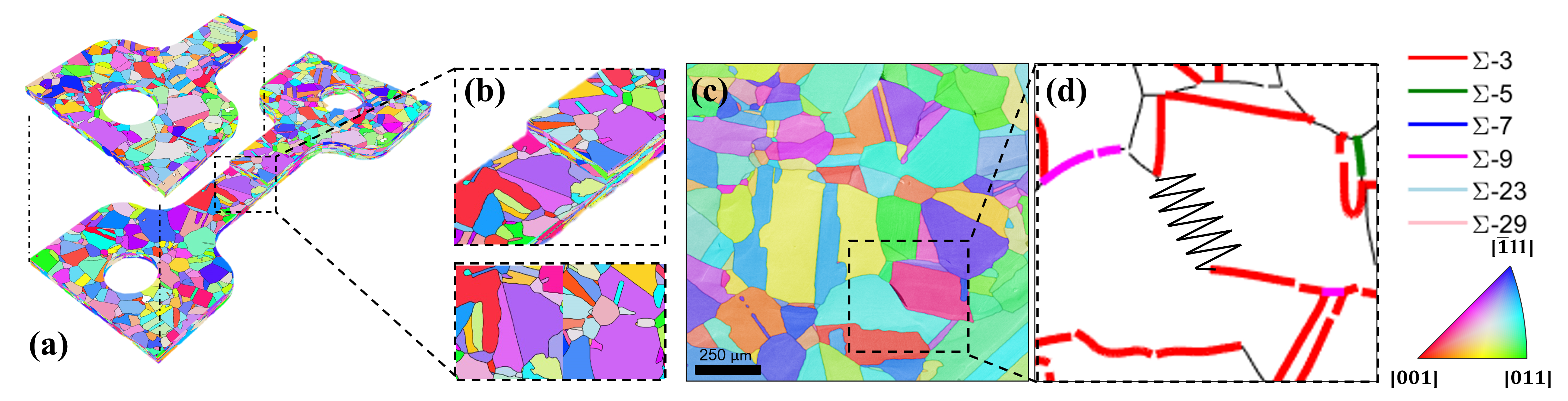}  % Adjust width as needed
    \caption{Surface and bulk intergranular cracking analysis: (a) material removal to expose the bulk, (b) IPF maps of surface and bulk, (c) IPF map of bulk material with a crack, and (d) GB characterization around the crack (depicted by a zigzag line). }
    \label{fig:Material removal}
\end{figure}

All cracks on the surface and bulk material were analyzed by scanning electron microscopy (SEM) and EBSD. Surface cracks were assessed using EBSD scans taken before the specimens were deformed, while for bulk cracks new EBSD maps were performed after fracture. There are roughly 7 layers of grains throughout the sample considering its thickness and average grain size. To characterize the bulk condition, the surface located halfway through the thickness of the sample was used. The bulk material was exposed by grinding to half thickness after the tensile test, followed by polishing with diamond suspensions to 0.25 \SI{}{\micro\meter} and finished with a 0.25 \SI{}{\micro\meter} OPS suspension for further EBSD characterization, as shown in Fig. \ref{fig:Material removal}. This procedure ensured that the GBs associated with bulk cracks were not the same as those of surface cracks, as sufficient thickness was removed to reveal a completely new surface of the material. 

\begin{figure}[th]
    \centering
    % First subfigure: Merged single length
    \begin{subfigure}[b]{0.5\textwidth}
        \centering
        \includegraphics[width=0.95\textwidth]{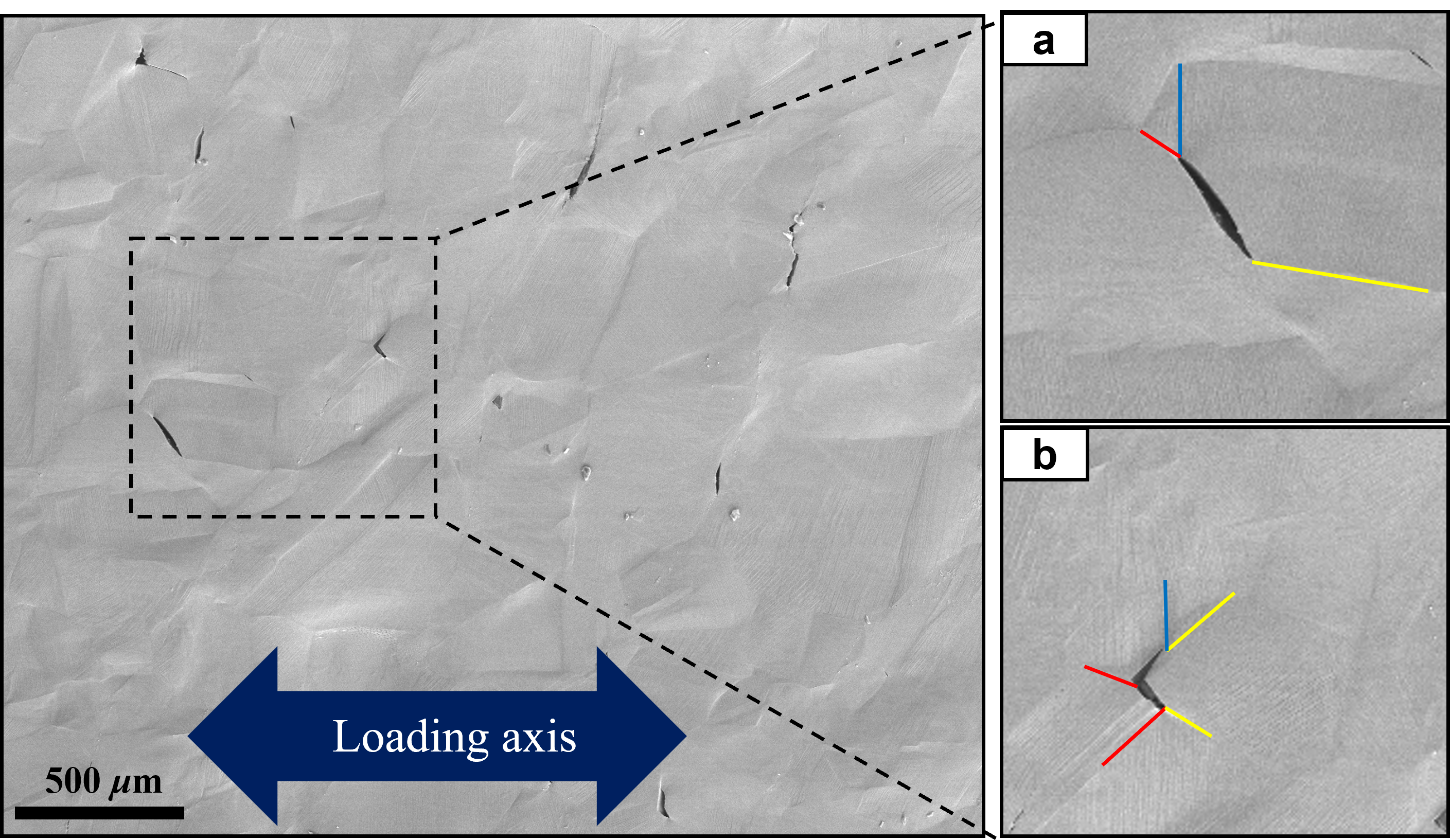}
    \end{subfigure}%
    % Second subfigure: Merged multiple length
    \begin{subfigure}[b]{0.5\textwidth}
        \centering
        \includegraphics[width=0.95\textwidth]{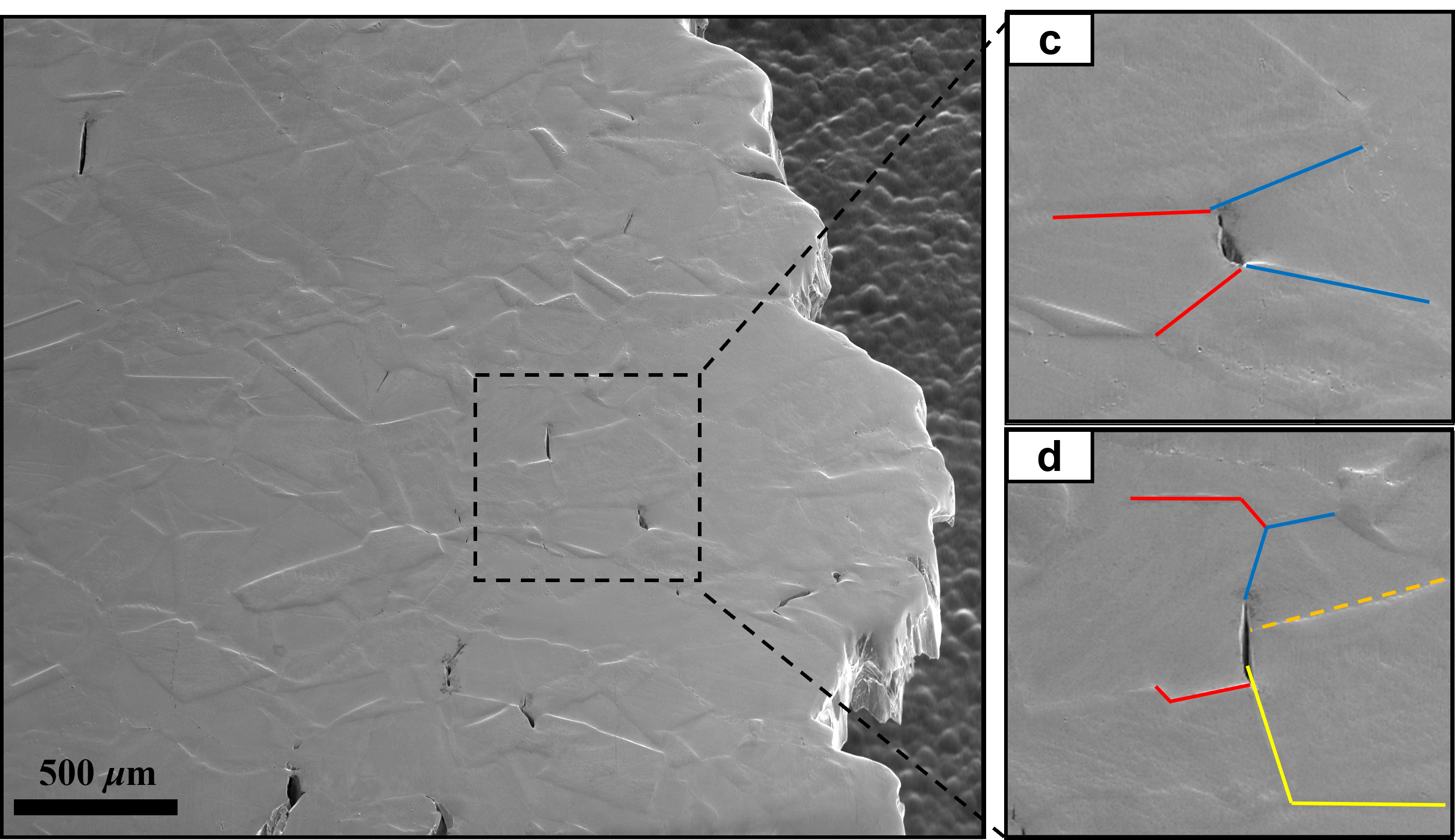}
    \end{subfigure}
    \caption{ The longitudinal surface of post-deformation specimens with highlighted cracks: (a) single-GB and (b) multi-GB surface cracks; (c) single-GB and (d) multi-GB bulk cracks after removing half of the material thickness.}
    \label{fig:Surface and bulk crack visualization}
    
\end{figure}

The analysis included quantifying the number of surface and bulk cracks using SEM and determining the fractions of $\Sigma$-3, other $\Sigma$-low and general GBs exhibiting intergranular cracking through SEM-EBSD characterization, according to \ref{appendix}. The resistance of GBs is discussed on the basis of comparing the fraction of fractured GBs (in terms of the number of GB segments) with the baseline microstructure. In addition, cracks were further classified as primary cracks if they are involved in the main crack leading to final failure, and secondary cracks if they occur in the GB but not in the main crack path. Cracks that contain more than one GB, involving crack initiation and propagation, are named multi-GB, and others passing through only one GB are referred to as single-GB cracks as represented in Fig. \ref{fig:Surface and bulk crack visualization}. Their length and inclination angles with respect to the loading axis were summarized to assess the extent to which plasticity (resolved shear stress at the grain boundaries) contributed to the fracture. To evaluate any contribution from electrochemical hydrogen charging, cracks at the surface of the material and at the bulk were compared.

All microscopy analyses to characterize the GBs and identify cracks were performed throughout the entire gauge length of the specimens using an Oxford Instruments EBSD system installed on a Zeiss Sigma V500 SEM.

\section{Results}
\label{sec:Results}
\subsection{Electrochemical charging and TDS measurement}
\label{sec:Electrochemical charging and TDS measurement}

The measured hydrogen concentration by TDS in the uncharged samples is 0.58 $\pm$ 0.12 wppm. The hydrogen concentrations after cathodic charging are presented in Table \ref{tab:hydrogen_concentration}, showing the effects of different electrolytes and current densities on hydrogen uptake in the samples. Two electrolytes were used: 0.6 \SI{}{\mole\per\liter} NaCl and 0.1 \SI{}{\mole\per\liter} NaOH, each tested at two different current densities, -4 \SI{}{\milli\ampere\per\centi\meter\squared} and -15 \SI{}{\milli\ampere\per\centi\meter\squared}. The measured potential of samples charged under the same conditions for TDS and tensile tests was identical. This consistent potential, to some extent, ensures a similar hydrogen evolution reaction and, consequently, comparable hydrogen concentrations between the TDS analysis and the tensile tested samples \cite{Livia2024}. The cathodic charging took around 40 minutes to stabilize, as indicated by a small change in potential at the initial stage. The variation in potential was within 0.03 \SI{}{\volt} during the subsequent stable cathodic charging.

\begin{table}[th]
\centering
\caption{Hydrogen concentration measurements for the various electrolytes, current densities, and potentials.}
\begin{tabular}{c c c c c}
\toprule
\textbf{Electrolyte} & \multicolumn{2}{c}{\textbf{0.6} \textbf{mol/L NaCl}} & \multicolumn{2}{c}{\textbf{0.1} \textbf{ mol/L NaOH}} \\
\cmidrule(lr){2-3} \cmidrule(lr){4-5}
\textbf{Current density (mA/cm$^2$)} & \text{-4} & \text{-15} & \text{-4} & \text{-15} \\
\textbf{Potential versus Ag/AgCl (V)} &  \text{-1.20} &  \text{-1.28} & \text{-1.30} &  \text{-1.50}\\
\cmidrule(lr){2-3} \cmidrule(lr){4-5}
\textbf{H concentration (wppm)} & \text{3.91} $\pm$ \text{0.35} & \text{3.45} $\pm$ \text{0.03} & \text{8.18} $\pm$ \text{0.34} & \text{14.47} $\pm$ \text{1.07} \\
\bottomrule
\end{tabular}

\label{tab:hydrogen_concentration}
\end{table}

No obvious surface damage was observed after cathodic charging up to a hydrogen concentration of 14 wppm achieved with 0.1 \SI{}{\mole\per\liter} NaOH at -15 \SI{}{\milli\ampere\per\centi\meter\squared}. The measured hydrogen concentration remained constant at approximately 4 wppm when using 0.6 \SI{}{\mole\per\liter} NaCl solution, even as the applied current density ranged from -4 \SI{}{\milli\ampere\per\centi\meter\squared} to -15 \SI{}{\milli\ampere\per\centi\meter\squared}. In contrast, charging with the 0.1 \SI{}{\mole\per\liter} NaOH solution led to higher hydrogen concentrations of 8 and 14 wppm at the same applied current densities.

Since no significant differences in hydrogen concentration were observed between NaCl charged samples, the charging conditions in this study effectively produced three different levels of hydrogen concentration: 4, 8 and 14 wppm. There was only a small variation between the two measurements for each charging condition, particularly under mild cathodic charging, such as when using NaCl or applying -4 \SI{}{\milli\ampere\per\centi\meter\squared}. The following sections show how these studied hydrogen concentrations were sufficient to induce significant HE in the nickel samples and discuss the effects of electrochemical charging on the susceptibility of GBs to hydrogen-assisted cracking. 

\subsection{Tensile Tests}
\label{sec:Tensile Tests}

\begin{figure}[th!]
    \centering
    \includegraphics[width=\textwidth]{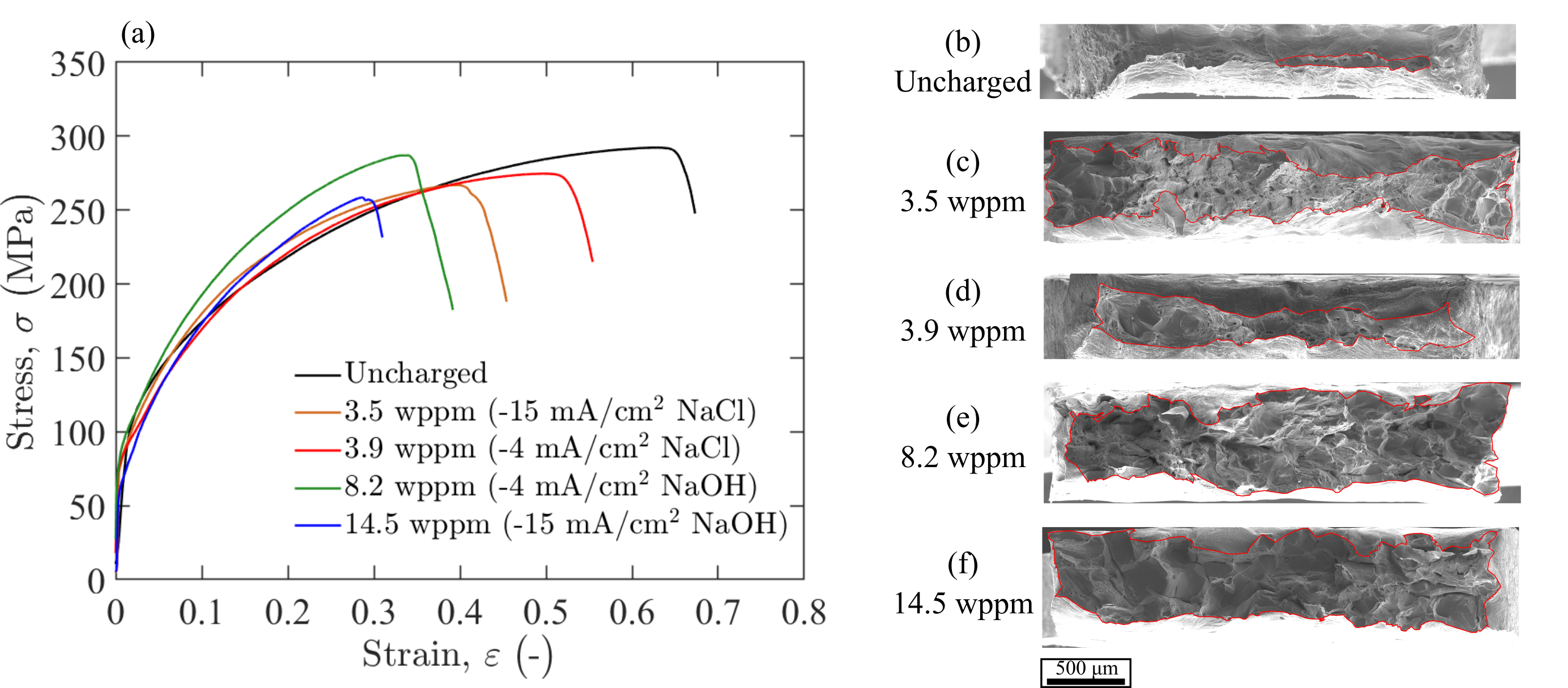}
    \caption{Engineering stress-strain curves (left) and fracture surfaces (right) of uncharged specimen and specimens with hydrogen concentration of 3.5, 3.9, 8.2 and 14.5 parts per million in weight (wppm).}
    \label{fig:Response cruve and fracture surface.}
\end{figure}

\begin{figure}[h]
    \centering
    \includegraphics[width=\textwidth]{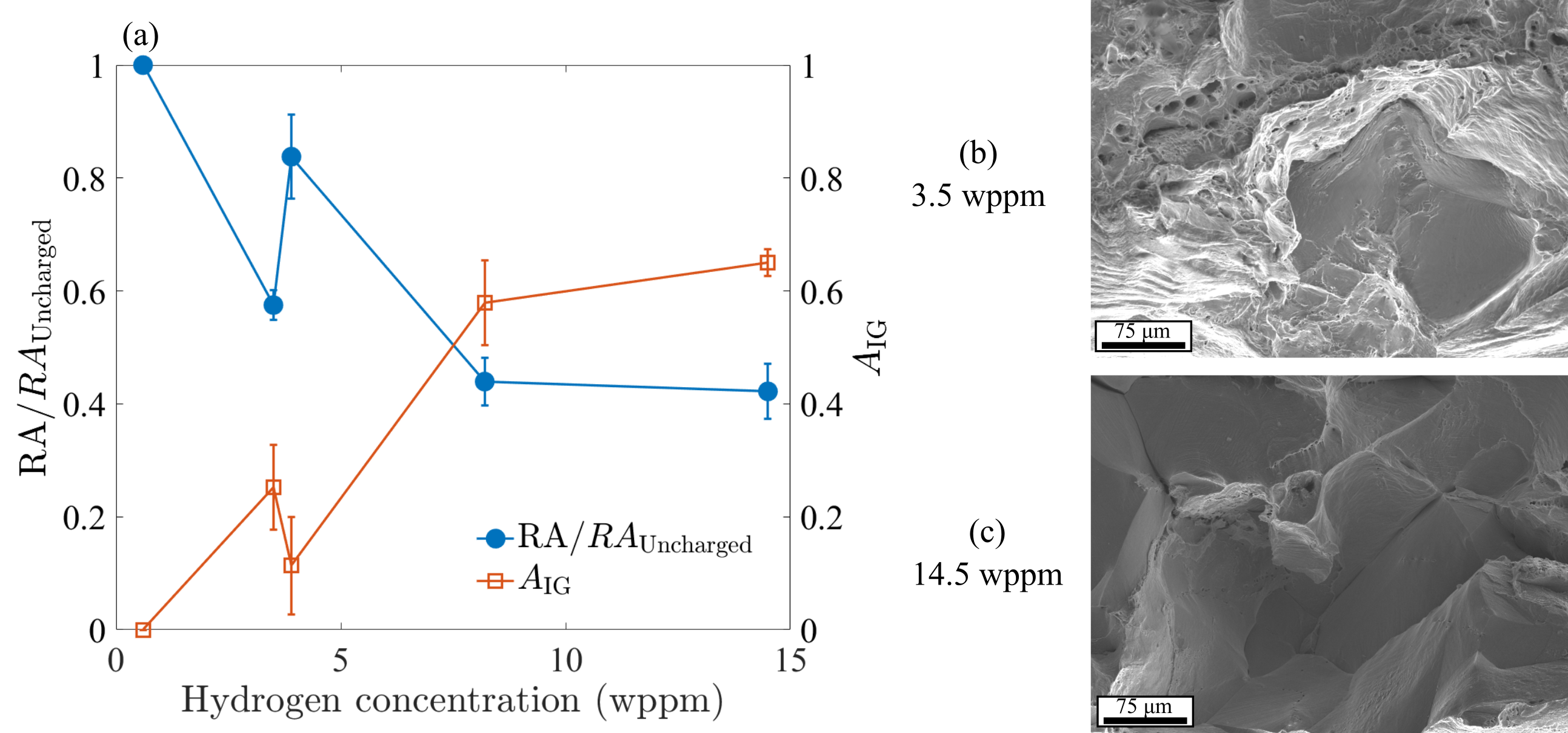};
    \caption{Quantifying HE in pure Ni: (a) Reduction in area, \(RA/RA_{\text{Uncharged}}\), and intergranular area, \(A_{\text{IG}}\), as functions of hydrogen concentration, indicating increased embrittlement with higher hydrogen concentration. Fracture surfaces for (b) 3.5 wppm with a mix of microvoids and intergranular fracture, whereas for (c) 14.5 wppm, a larger fraction of intergranular fracture corresponds to a higher embrittlement level.}
    \label{fig:Fractography_RA_IG}
\end{figure}

Tensile tests were performed with specimens charged in all the conditions displayed in Table \ref{tab:hydrogen_concentration} using a strain rate of $2 \times 10^{-4}$ \SI{}{\second^{-1}}, as shown in Fig. \ref{fig:Response cruve and fracture surface.}. All specimens exhibited HE, evidenced by a reduction in the fracture strain ($EI\textsubscript{FS}$) relative to the uncharged condition, while no significant changes in work-hardening behaviour were observed. HE is also evaluated based on the reduction in area ($EI\textsubscript{RA}$, RA\(/RA_{\text{Uncharged}}\)), and the fraction of intergranular fracture ($A_{\text{IG}}$) on the fracture surface, as can be seen in Fig. \ref{fig:Fractography_RA_IG}(a).

As expected, embrittlement became more pronounced with increasing hydrogen concentration. Although samples charged with 3.5 and 3.9 wppm showed similar trends in $A\textsubscript{IG}$, notable differences in $EI\textsubscript{FS}$ and $EI\textsubscript{RA}$ can be observed. A marked increase in $A\textsubscript{IG}$ was evident between 4 and 8 wppm. However, beyond 8 wppm, the embrittlement level is plateauing, with no further significant changes in $EI\textsubscript{FS}$ or $EI\textsubscript{RA}$, yet showing a 10\% increase in $A_{\text{IG}}$. This indicates that a comprehensive assessment of the severity of embrittlement is best achieved by considering the three parameters $EI\textsubscript{FS}$, $EI\textsubscript{RA}$, and $A\textsubscript{IG}$ in combination. The trend shown in Fig. \ref{fig:Fractography_RA_IG}(a) agrees with previous observations of nonlinear relationships between embrittlement severity and hydrogen concentration, with initially sharp reductions in toughness followed by a plateau at a concentration-invariant toughness \cite{Bechtle2009}. The small differences in HE severity between specimens with 8 and 14 wppm indicate that 14 wppm is probably close to the plateau where the maximum embrittlement level for nickel is reached. On the other hand, the dispersion of the data between 3.5 and 3.9 wppm is likely because, at these hydrogen levels, small differences in concentration between the tensile and TDS samples lead to a larger scatter in HE.

The uncharged specimen's fracture surface exhibited extensive evidence of ductile microvoid coalescence and a complete absence of intergranular faceting, consistent with the observed RA of 96\%. In contrast, the fracture surfaces of all four cathodic charging conditions show widespread intergranular fracture which increased in area fraction as hydrogen content increased. Such results are supported by the smaller reduction in area observed for each condition relative to the uncharged baseline. Critically, intergranular fracture was consistently observed in the specimen mid-thickness in all hydrogen-charged cases, despite the presence of mixed failure modes, including intergranular fracture and microvoid growth for the 3.5 and 3.9 wppm (Fig. \ref{fig:Fractography_RA_IG}(b)). This mixed failure mode is, to a certain extent, also observed in the sample with 14 wppm where the percentage of intergranular fracture was 65\% (Fig. \ref{fig:Fractography_RA_IG}(c)). In general, HE indexes ($EI\textsubscript{FS}$, $EI\textsubscript{RA}$, and $A\textsubscript{IG}$) together with the intergranular fracture distribution on the fracture surfaces, especially at 14 wppm, not only confirm that the hydrogen concentration was sufficient to induce significant intergranular fracture, but also suggest a uniform hydrogen distribution within the samples.

\subsection{Hydrogen-assisted intergranular cracking}

\begin{figure}[h]
    \centering
    \includegraphics[width=\textwidth]{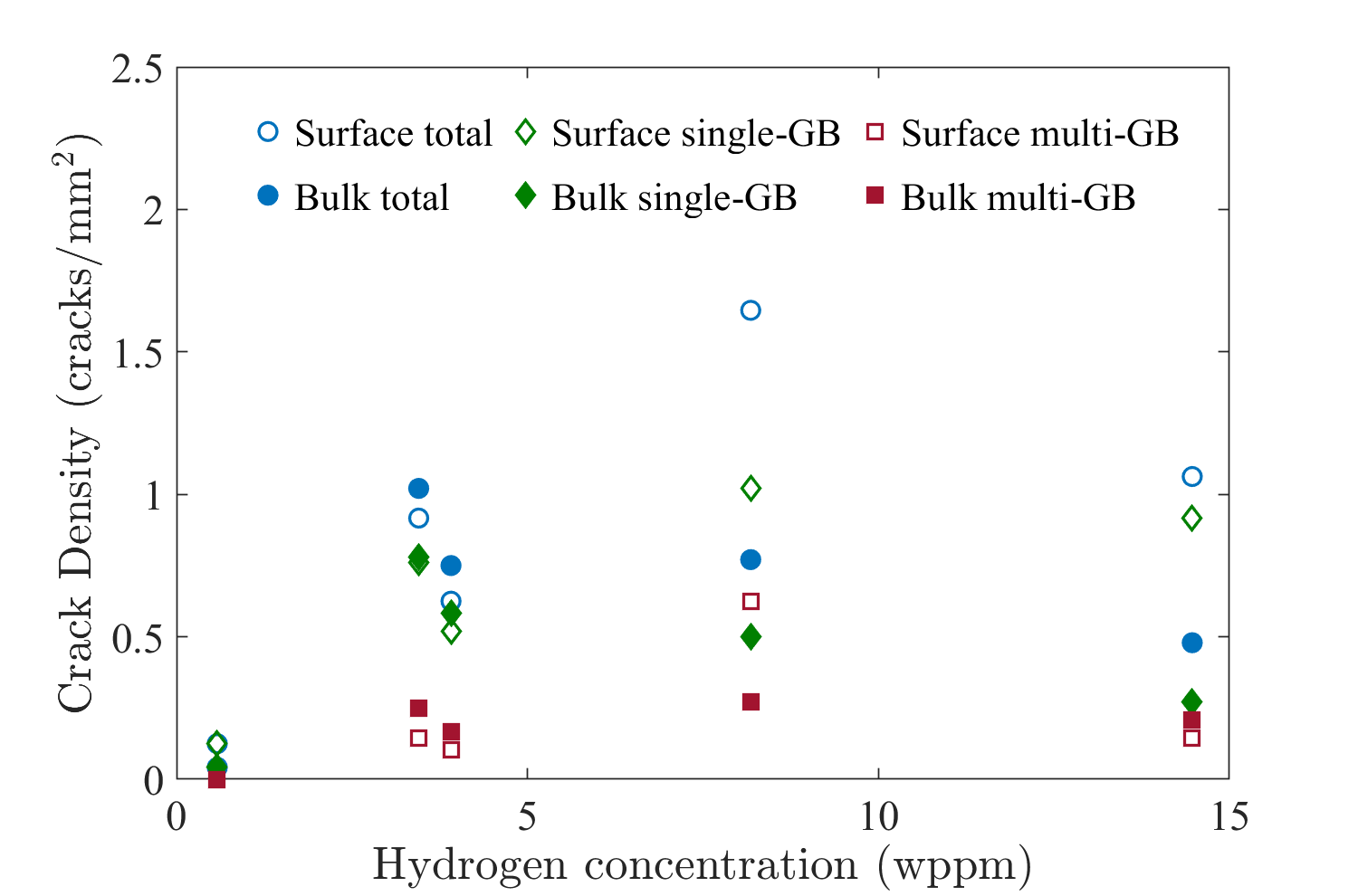};
    \caption{GB crack densities: total, single-GB and multi-GB crack densities for the surface (hollow symbols) and bulk (solid symbols), considering the average number of cracks observed in the gauge region of two samples for each hydrogen concentration. The crack densities for the uncharged sample are included for comparison.}
    \label{fig:Number of cracks.}
\end{figure}

The number density of intergranular cracks (number of cracks per gauge length surface area, which is 8 mm x 3 mm $=$ 24 mm$^2$) is summarized in Fig. \ref{fig:Number of cracks.}. Post-deformation intergranular cracks were rarely observed in the uncharged specimen, with only three surface cracks and one bulk crack detected. On the other hand, there was an average of 15 to 40 surface and 12 to 25 bulk intergranular cracks in the hydrogen-charged specimens, depending on the charging conditions. For the samples with 3.5 and 3.9 wppm, there is a similar number of cracks on the surface and in the bulk material, suggesting that these charging conditions did not induce surface damage. In contrast, there is a substantially larger number of cracks on the surface compared to the bulk material for the samples charged with 8 and 14 wppm. This indicates that more intense hydrogen evolution and/or increased hydrogen absorption during these cathodic charging conditions may facilitate the formation of surface cracks. It is also worth noting that the higher deformation levels attained for the low hydrogen content cases will result in higher hydrostatic stresses at the center of the samples, which could drag hydrogen from the surface and further contribute to higher predominance of bulk cracks (versus the lower necking, high concentration tests). 

\begin{figure}[H]
    \centering
    \includegraphics[width=\linewidth]{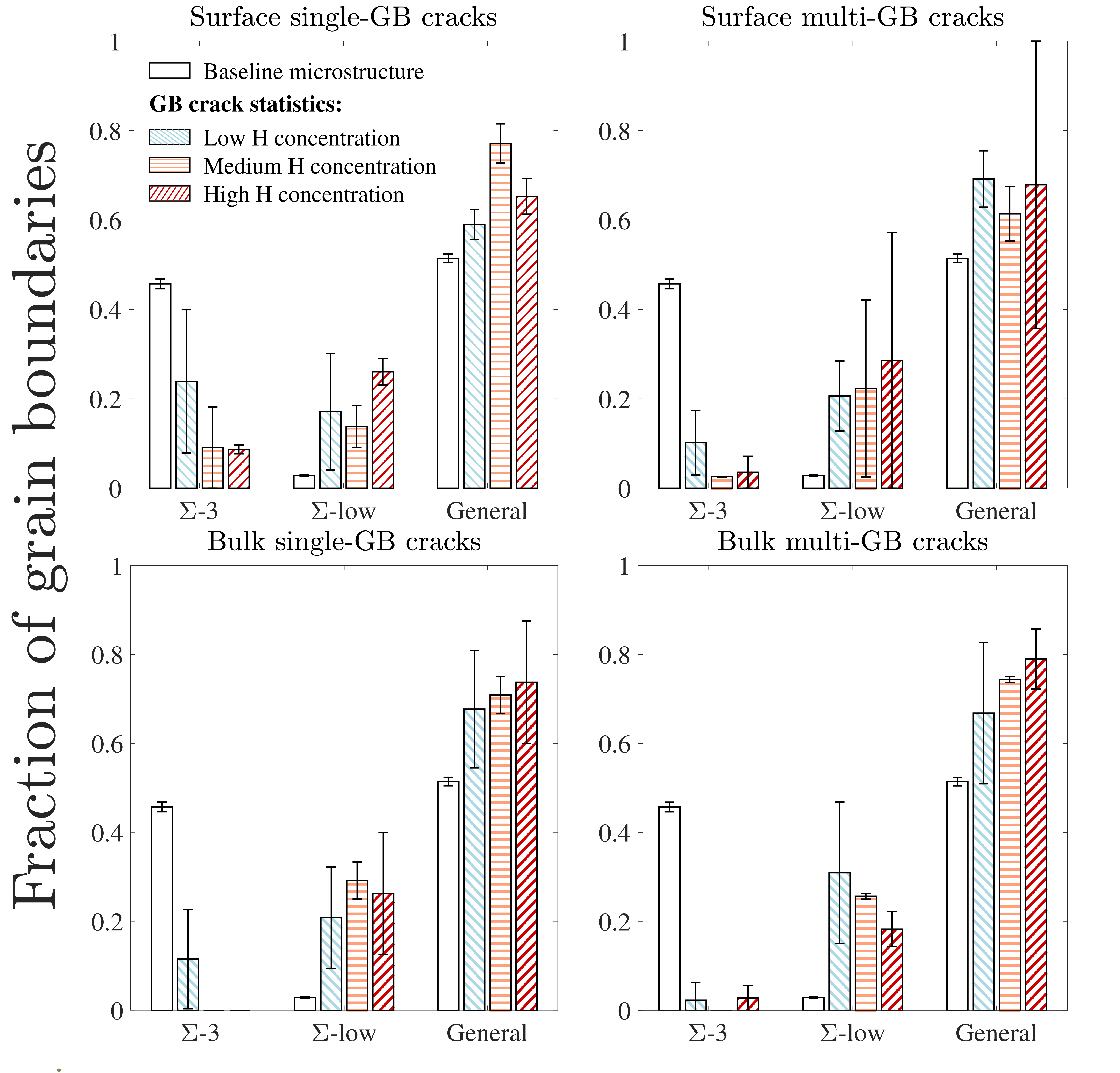}
    \caption{GB crack statistics compared to the baseline microstructure. The fractions of fractured GBs are shown according to their $\Sigma$ values, on the surface or in the bulk, involved in single-GB or multi-GB secondary cracks for low (blue), medium (orange), and high (red) hydrogen concentrations. Error bars represent the standard deviation based on all samples for each concentration. The groups of GBs are based on their CSL values, $\Sigma$-3, $\Sigma$-low (3 < $\Sigma \leq 29$) and general GBs.}
    \label{fig:SecondaryCrackFraction}
\end{figure}

\begin{figure}[H]
    \centering
    % First row
    \includegraphics[width=0.6\textwidth]{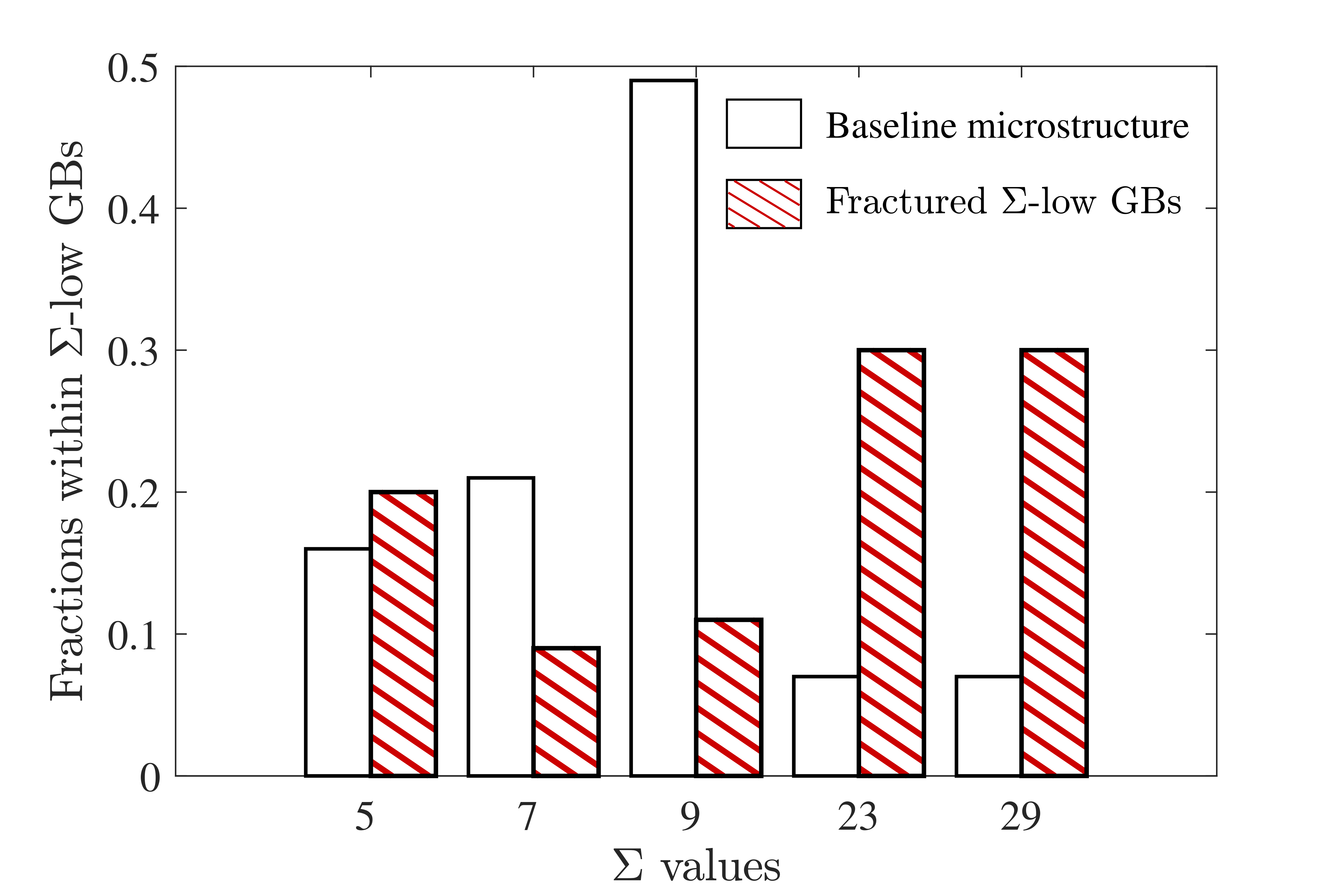}
    \caption{GB crack susceptibility within the $\Sigma$-low category: the distribution of all fractured $\Sigma$-low GBs (red diagonal stripes) is compared to the baseline $\Sigma$-low distribution (white). Although all belong to the same classification, $\Sigma$-23 and $\Sigma$-29 GBs exhibit higher susceptibility to hydrogen-assisted fracture.}
    \label{fig:FractionWithin5to29}
\end{figure}

\begin{figure}[H]
    \centering

    \includegraphics[width=\textwidth]{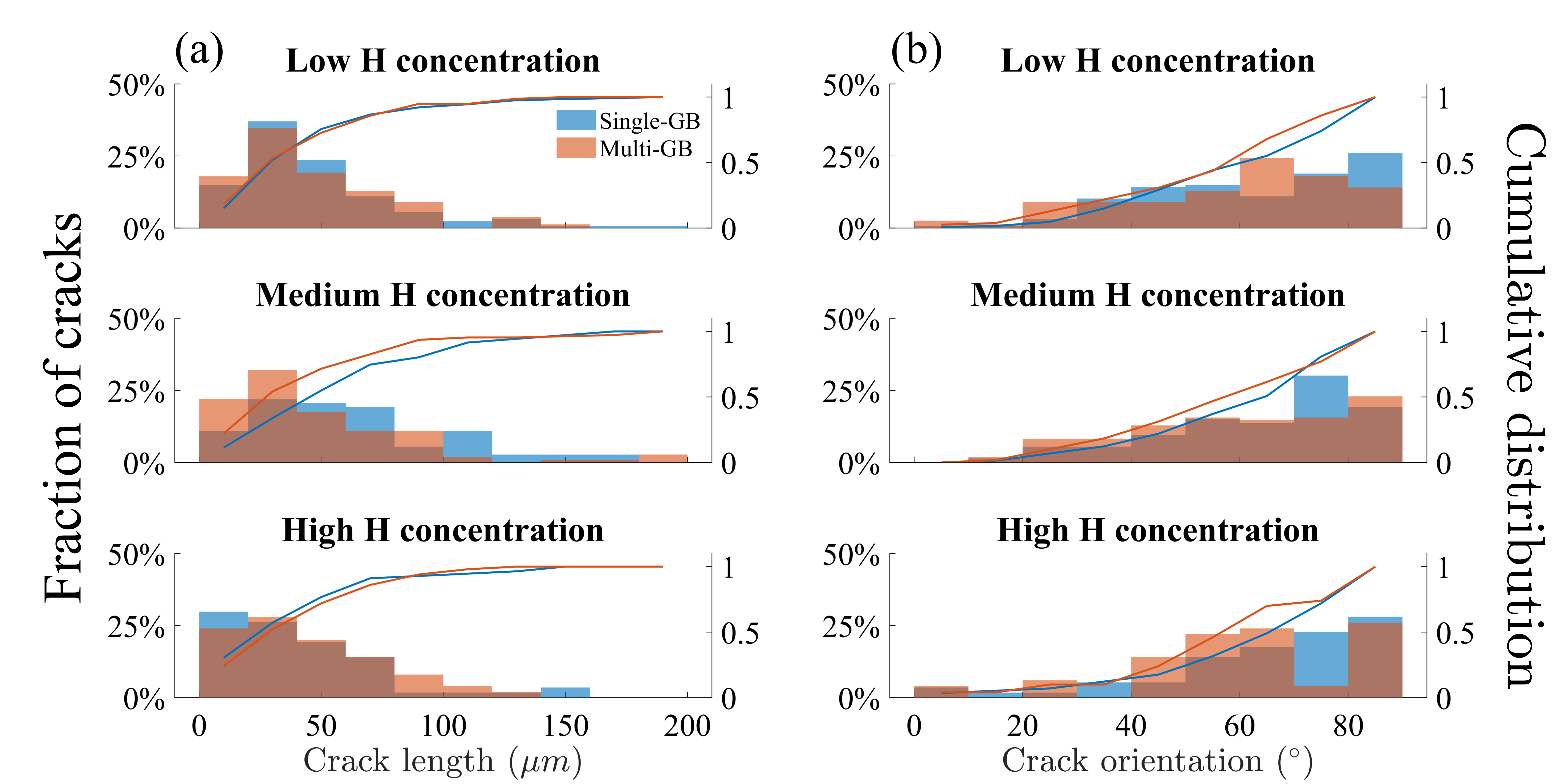}

    \caption{Geometric properties of the identified secondary cracks: (a) length and (b) orientation angle distributions. The solid lines represent the cumulative distributions. }
    \label{fig:crack_distribution_comparison}
\end{figure}

\begin{figure}[H]
    \centering
    % First row
    \includegraphics[width=\textwidth]{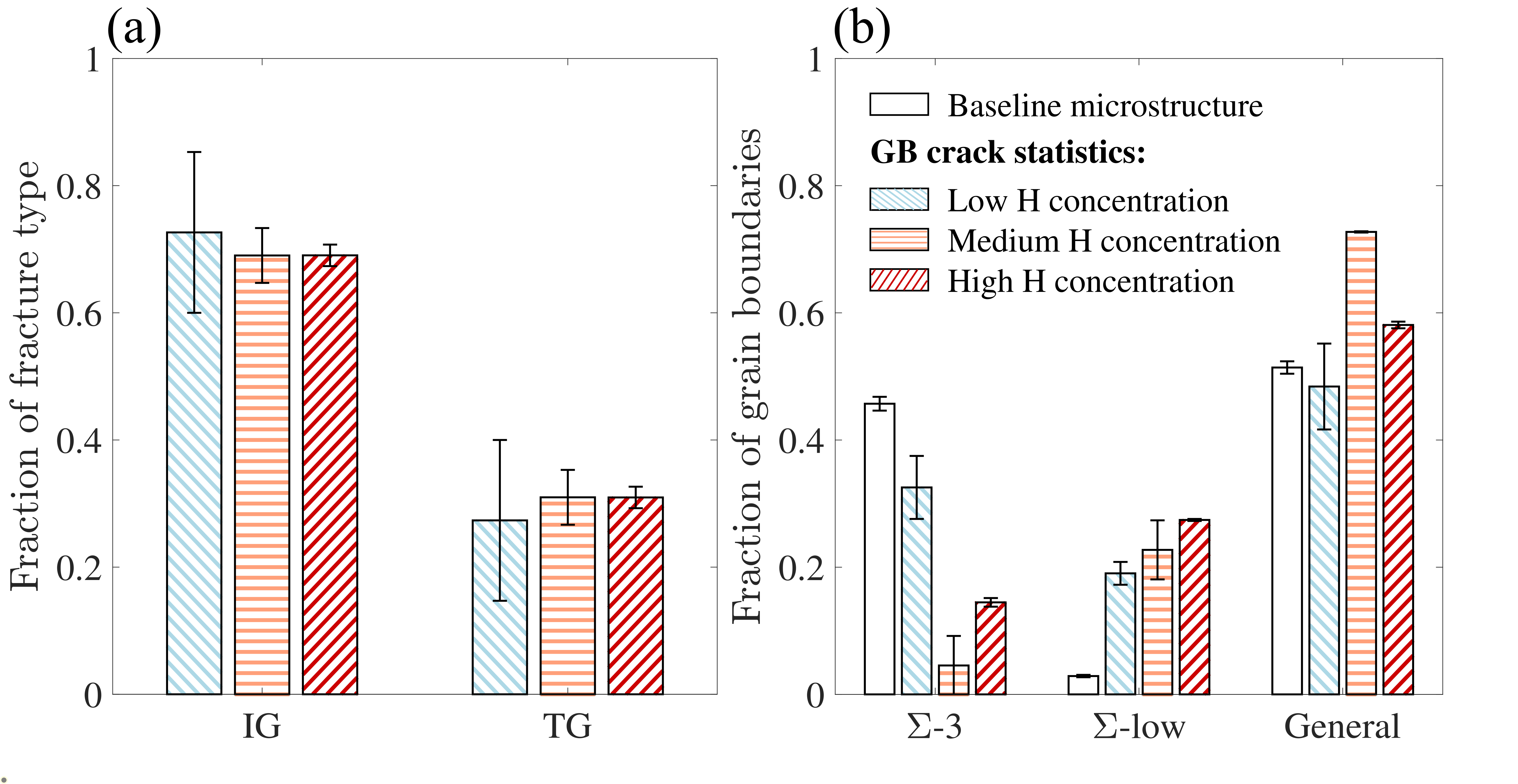}
    \caption{Fractions of (a) intergranular, IG, and transgranular, TG, in the primary cracks, and (b) GB crack statistics for the IG cracks compared to the baseline microstructure. Error bars represent the standard deviation based on all samples for each concentration.}
    \label{fig:Primary crack grain boundary type summary}
\end{figure}

\begin{figure}[H]
    \centering
    % First row
    \centering
    \includegraphics[width=\textwidth]{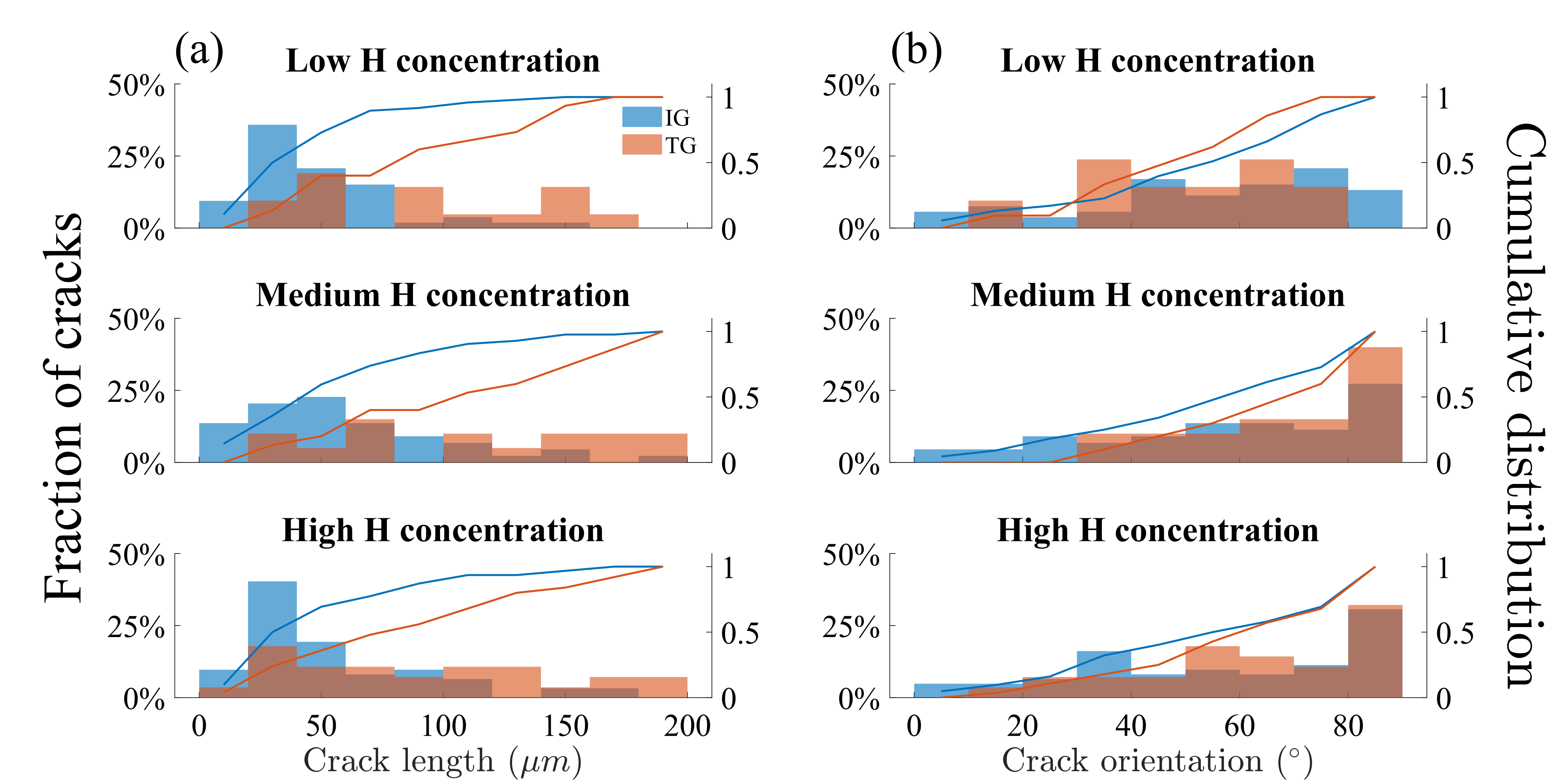}
    \caption{Geometric properties of the identified primary cracks: (a) length and (b) orientation angle distributions. The solid lines represent the cumulative distributions.}
    \label{fig:Primary crack length and orientation summary}
\end{figure}

The fractions of $\Sigma$-3, $\Sigma$-low and general GBs are plotted according to their locations, crack categories, and hydrogen concentrations in Fig. \ref{fig:SecondaryCrackFraction}. The samples charged with 3.5 and 3.9 wppm were grouped for clarity due to the similarity in observed HE levels, fracture surface morphologies, and crack density. Thus, in Fig. \ref{fig:SecondaryCrackFraction} to \ref{fig:Primary crack length and orientation summary}, hydrogen concentrations are classified as low (3.5-3.9 wppm), medium (8 wppm) and high (14 wppm). The GB baseline microstructure has an average distribution of 46 $\pm$ 1.1\% $\Sigma$-3, 3 $\pm$ 0.2\% $\Sigma$-low and 51 $\pm$ 1\% general GBs. These baseline fractions were obtained from $\sim5000$ GBs per specimen. The small variations in the fractions of GBs in the baseline microstructure between all the samples confirms that they had a nominally identical GB distribution before tensile testing. 

From the results in Fig. \ref{fig:SecondaryCrackFraction}, the relative fractions of fractured GBs appear to be independent of their location (surface or bulk) and crack type (single or multi-GB). The $\Sigma$-3 GBs consistently exhibit the lowest fraction among the fractured GBs, showing a substantial decrease compared to the baseline, regardless of location, crack type, or hydrogen concentration. Specifically, $\Sigma$-3 GBs accounted for 8\%, $\Sigma$-low GBs for 24\%, and general GBs for 68\% of all secondary cracks observed. In addition, the occurrence of fractured $\Sigma$-3 GBs decreases with increasing hydrogen concentrations. These findings thus demonstrate that $\Sigma$-3 GBs have the highest resistance to hydrogen-assisted cracking at all hydrogen concentrations. This conclusion applies to both crack initiation and propagation, as the fraction of fractured GBs shows no significant difference between single-GB and multi-GB cracks, indicating that susceptibility is independent of the number of GBs involved.

Although there is a difference in the number of cracks between the surface and bulk (Fig. \ref{fig:Number of cracks.}), they have similar relative proportions of the types of fractured GBs. This suggests that while cracking may be enhanced by the intensity of cathodic charging, this effect does not alter which types of GBs are more or less susceptible to hydrogen-assisted intergranular cracking. Finally, Fig. \ref{fig:SecondaryCrackFraction} also reveals that $\Sigma$-low GBs (3 < $\Sigma \leq 29$) have higher post-deformation fracture fractions compared to the baseline microstructure. A detailed analysis (Fig. \ref{fig:FractionWithin5to29}) shows that this is due to the greater susceptibility of $\Sigma$-23 and $\Sigma$-29, while GBs with lower $\Sigma$ values (especially  $\Sigma$-7 and $\Sigma$-9) demonstrate resistance to hydrogen-assisted cracking with smaller fractions of fractured GBs than the baseline. This figure combines the fractured GBs within the $\Sigma$-low range from all categories (surface and bulk, single-GB and multi-GB) and hydrogen concentrations (low, medium and high).

A summary of the crack lengths and orientation angles of the GBs exhibiting intergranular cracking is presented in Fig. \ref{fig:crack_distribution_comparison}. Surface and bulk cracks are combined in this analysis. In Fig. \ref{fig:crack_distribution_comparison}(b), a crack orientation angle of 0\SI{}{\degree} indicates alignment parallel to the loading axis, while 90\SI{}{\degree} corresponds to perpendicular alignment. For multi-GB cracks, the angles of all individual GB crack segments are taken into account, i.e. if a multi-GB crack spans two grain boundaries, two crack segment angles are considered. At all hydrogen concentrations, GBs with orientation angles below 20\SI{}{\degree} are rare. Cracks exhibit preferential orientation, with a higher frequency of angles between 70\SI{}{\degree} and 90\SI{}{\degree}, which can be seen to some extent in Fig. \ref{fig:Surface and bulk crack visualization}. Furthermore, hydrogen concentration does not appear to influence crack length distributions for either single-GB or multi-GB cracks, with typical lengths ranging from 20 to 70 \SI{}{\micro\meter} for both.

GB fraction and crack geometry analyses were also performed on primary cracks (i.e., cracks responsible for the final failure) by tracking their paths on the longitudinal view SEM images of the fractured samples and comparing them with the EBSD maps of the undeformed samples. The primary cracks exhibit a combination of intergranular (IG) and transgranular (TG) fractures. The fractions of these two failure modes and of GBs in the IG cracks are shown in Fig. \ref{fig:Primary crack grain boundary type summary}. There are around 70\% IG and 30\% TG for all the studied hydrogen concentrations. Similar to secondary cracks, the $\Sigma$-3 GBs consistently show the highest resistance to hydrogen-assisted intergranular cracking. The IG crack length for primary cracks is comparable to that of secondary cracks, generally ranging from 20 to 70 \SI{}{\micro\meter} (Fig. \ref{fig:Primary crack length and orientation summary}(a)), while TG cracks exhibit greater variability in length. The orientation angles of the fractured GBs show a more pronounced trend around 80\SI{}{\degree} and 90\SI{}{\degree} (Fig. \ref{fig:Primary crack length and orientation summary}(b)), especially for medium and high hydrogen concentrations, reinforcing the contribution of decohesion to the cracking of these interfaces. 

\section{Discussion}

By analyzing 349 secondary and 159 primary intergranular cracks, this work provides strong evidence that $\Sigma$-3 GBs are more resistant to hydrogen-assisted cracking. Furthermore, the lower susceptibility to HE of low-$\Sigma$ GBs and the orientation angles of the fractured GBs relative to the loading axis suggest that failure is predominantly driven by decohesion rather than plasticity.

\subsection{Embrittlement and density of IG cracks}

Firstly, the findings of the present work appear to be universal and not linked to the hydrogen charging approach adopted. The increasing levels of embrittlement with hydrogen concentration after electrochemical charging, linked to higher fractions of intergranular fracture, were similar to previous findings for a similar Ni material charged with hydrogen gas at high temperature and pressure \cite{Bechtle2009}. This embrittlement was also accompanied by a decrease in the density of secondary integranular cracks observed at the bulk of the specimens with the highest hydrogen concentration. This can be explained by the expected decrease in fracture toughness with hydrogen concentration, enabling fracture to occur with an increasingly lower level of internal damage, which in this work consists mainly of intergranular cracks. Furthermore, a similar crack density is observed at the surface and bulk for the smallest tested hydrogen concentrations ($\approx$ 4 wppm). On the other hand, the surface is about twice as susceptible to cracking than the bulk for the tests with 8 and 14 wppm, likely due to the more severe applied cathodic charging, although no obvious damage to the surface (e.g., blisters \cite{Lu2019}, slip lines and cracks \cite{Pérez2011}) was observed after charging.

\subsection{Mechanism of hydrogen-assisted IG cracking in nickel}

The current results demonstrate that the $\Sigma$ value of a GB correlates well with its likely susceptibility to hydrogen-assisted cracking. Although the $\Sigma$ value alone does not fully characterize the crystallography of GBs \cite{Hanson2018}, the present data show that general GBs exhibit increased susceptibility, while a lower susceptibility is observed for $\Sigma \leq 9$ boundaries. This result is in agreement with previous experimental works (e.g., \cite{Bechtle2009, Kwon2018}) that suggest that increasing the fraction of low-$\Sigma$ boundaries through grain boundary engineering methods could be an effective strategy to mitigate HE. Moreover, this study provides a more comprehensive statistical demonstration of the generality of this susceptibility trend ($\Sigma$-high vs $\Sigma$-low) for Ni over a wide range of hydrogen concentrations and whether the cracks are bulk or surface, primary or secondary, single or multi-GB, even under conditions where the metal surfaces appear to be damaged by severe cathodic charging ($\geq$ 8 wppm). 

\begin{figure}[H]
    \centering
    % First row
    \includegraphics[width=\textwidth]{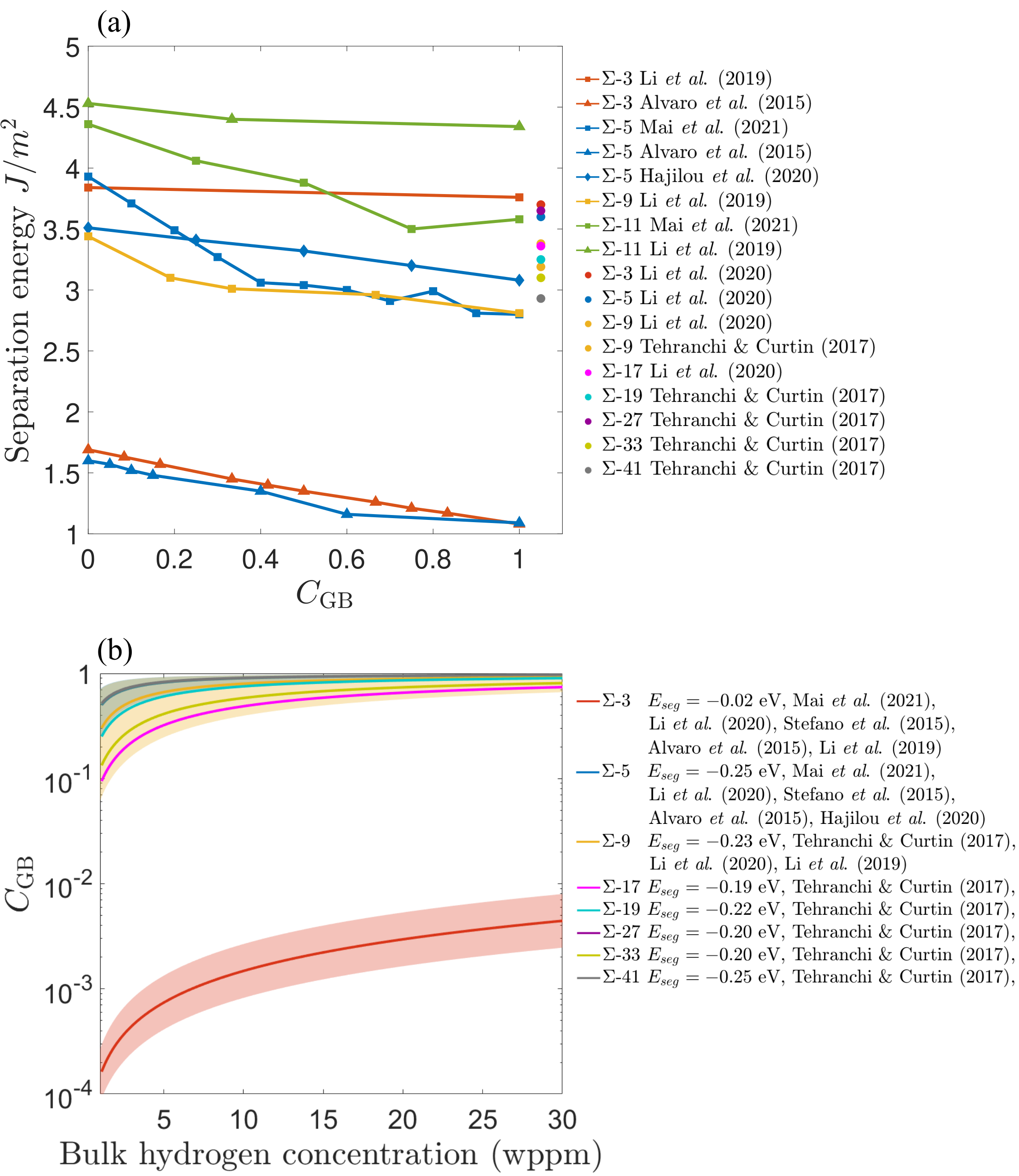}
    \caption{(a) GB separation energy as a function of hydrogen coverage from DFT and MD simulations \cite{Li2019, Mai2021, Hajilou2020, Alvaro2015}. The circles indicate conditions under which near full hydrogen coverage at the GB is achieved for a given bulk atomic concentration of 0.001 (17 wppm) \cite{Tehranchi2017-II, Li2020DFT}. (b) Hydrogen coverage at grain boundaries as a function of bulk hydrogen concentration. The shaded areas represent variations in coverage due to different segregation energies reported in the literature \cite{Tehranchi2017-II, Mai2021, Li2020DFT, DiStefanoDavide2015, Alvaro2015,  Li2019,  Hajilou2020}. The mean segregation energy values considered are given in the legend. For works where multiple values are provided, the highest segregation energy is used for each GB type. 
    }
    \label{fig:Atomistic summary}
\end{figure}

The observed link between hydrogen-assisted intergranular cracking with GB character can be explained based on thermodynamics and kinetics considerations of decohesion. By defining the density of coincident lattice sites, the $\Sigma$ value serves as a proxy for the atomic structure of the GB, which directly influences the grain boundary energy, as well as the hydrogen segregation energy and the fracture energy as a function of hydrogen occupancy. Atomistic studies have predicted linear reductions of cohesive energies of GBs with hydrogen occupancy (\(C_{GB}\)) depending on their $\Sigma$ values, as summarized from multiple studies in Fig. \ref{fig:Atomistic summary}(a). However, it is unclear which grain boundary type experiences the least reduction in separation energy with increasing hydrogen coverage due to substantial discrepancies in atomistic simulation results reported in the literature. 

On the other hand, calculations and experimental results corroborate the higher resistance of $\Sigma$-3 due to the significantly less favorable trapping \cite{DiStefanoDavide2015, Alvaro2015, Ma2021}. The hydrogen coverage at grain boundaries (\(C_{GB}\)) as a function of the hydrogen bulk concentration and their segregation energies is shown in Fig. \ref{fig:Atomistic summary}(b). These values were calculated according to the Langmuir-McLean isotherm \cite{Lassila1988, Fernndez-Sousa2020} \begin{equation}
\frac{C_{GB}}{1 - C_{GB}} = \frac{C_{Bulk}}{1 - C_{Bulk}} \exp\left(\frac{-E_{seg}}{kT}\right)
\label{eq:McLean}
\end{equation} where \(C_{Bulk}\) is the atomic concentration of hydrogen in the bulk, \(E_{Seg}\) is the segregation energy for hydrogen to grain boundary sites, $k$ is the Boltzmann's constant and $T$ is the temperature in Kelvin. In Fig. \ref{fig:Atomistic summary}(b), \(C_{Bulk}\) is converted into wppm to facilitate comparison with the values in this work (Table \ref{tab:hydrogen_concentration}). The segregation energies reported in the literature used to calculate \(C_{GB}\) are also provided. For the GB types where \(E_{Seg}\) is reported in multiple studies, the average value is taken and a shaded area is used to represent the scatter considering the standard deviation of the values. It is possible to see from Fig. \ref{fig:Atomistic summary}(b) that the $\Sigma$-3 boundary exhibits the least favorable hydrogen segregation energy, with a mean value of -0.02 \SI{}{\electronvolt} and marginal scattering (-0.01 to -0.04 \SI{}{\electronvolt} \cite{Alvaro2015, Li2020DFT, Li2019, Mai2021, DiStefanoDavide2015}). Therefore, increasing hydrogen coverage at the $\Sigma$-3 boundary in pure nickel is particularly difficult, greatly limiting hydrogen-assisted intergranular cracking to happen.

Recent studies have shown that the thermodynamically predicted equilibrium segregation and fracture energy decreases of hydrogenated boundaries alone could not explain cleavage along GBs in Ni \cite{Tehranchi2017-II, Tehranchi2020}. Hydrogen-enhanced decohesion of GBs could only be predicted in the atomistic simulations when it was included kinetics considerations related to hydrogen accumulation at atomic-scale crack tips and diffusion towards fractured surfaces \cite{Tehranchi2020}. The decrease in local Mode I crack tip stress critical intensity factor ($K_{\text{Ic}}$) resulting from these phenomena is also dependent on the GB's crystallography, and could hence be used to explain the observed differences in HE resistance according to the GB $\Sigma$ value. Therefore, the results of this work are in line with the decohesion mechanism in which the low-energy GBs, especially $\Sigma$-3, are expected to be more resistant to cracking in the presence of a given hydrogen concentration due to their lesser segregation and greater cohesive strength. 

\subsection{Difference between cracking susceptibility for Ni and Ni alloys}

In contrast to what is predicted by thermodynamic-kinetic considerations of decohesion and observed in the present work, recent studies with Ni alloys have shown greater susceptibility of the $\Sigma$-3 coherent twin GBs to hydrogen-assisted cracking nucleation \cite{Seita2015, Liu2024, Zhang2020}. These results were explained based on the localized plasticity activity at these GBs. The fact that in the present work no such plasticity-enhanced cracking nucleation at $\Sigma$-3 boundaries is observed suggests that this cracking mechanism is less favored in pure Ni than in Ni alloys. One possible explanation is that the susceptibility to hydrogen-assisted cracking of twin boundaries in FCC materials depends on their stacking fault energy (SFE), with low SFE alloys being more susceptible due to enhanced planar slip and impediment of cutting through twin boundaries \cite{Chen2013}. This susceptibility may also be enhanced by the $\gamma'$ and/or $\gamma''$ precipitation observed in several Ni alloys, such as the Inconel 725 studied in \cite{Seita2015, Liu2024}. Therefore, pure nickel with a higher SFE than its alloys and without precipitation would, as observed in this work, be more resistant to twin boundary cracking enhanced by hydrogen and plasticity interactions. A final analysis of crack orientation was performed in this work to elucidate the role of resolved shear stress in GB cracking. Cracks oriented nearly normal to the loading axis are observed more frequently, corroborating the decohesion mechanism. Therefore, once factors that induce plasticity localization are removed (e.g., solutes and precipitates), as in this work using pure Ni, the results show that both the grain boundary character and the orientation relative to the load axis are key factors for the occurrence of hydrogen-assisted intergranular cracks.

\section*{Conclusion}
This study characterizes the susceptibility of nickel's surface and bulk GBs to hydrogen-assisted intergranular cracking under tensile loading across a wide concentration range (4 to 14 wppm) using SEM-EBSD. The main findings are as follows:

\begin{itemize}
    \item $\Sigma$-3 GBs consistently exhibit the highest resistance to hydrogen-assisted intergranular cracking. This resistance is reflected in crack initiation and propagation.
    \item Cathodic charging can promote hydrogen-assisted surface cracks. However, the relative susceptibility of the GBs remains unchanged.
    \item Embrittlement and intergranular fracture increase with hydrogen concentration. This is accompanied by a decrease in the bulk secondary cracking, corroborating that a lower degree of damage is required for failure to occur as the hydrogen concentration increases.
    \item The length and orientation of primary and secondary cracks show no sensitivity to hydrogen concentration. Most crack lengths are between 20 and 70 µm for an average grain size of 101 µm. Crack orientations >70$\degree$ (relative to the applied load) are observed more frequently.
    \item Intergranular cracking susceptibility in nickel can be explained by thermodynamic-kinetic considerations of hydrogen-enhanced decohesion. 
    \item Hydrogen-induced plasticity-assisted cracking at twin boundaries is less favored in pure Ni than in its alloys.  
\end{itemize}

Therefore, this work brings new insights into the intrinsic resistance of grain boundaries to hydrogen-assisted cracking. It contrasts with scenarios, such as that of certain Ni alloys, in which the local stresses from plastic localization can obfuscate GB susceptibility. Its findings lay the foundation for more holistic engineering strategies for Ni alloys, taking into account not only the GB character but also the complexity of the structural features within them. 

\section*{Declaration of Competing Interest}

The authors declare that they have no known competing financial interests or personal relationships that could have appeared to influence the work reported in this paper.

\section*{Acknowledgements}

The authors acknowledge financial support from the EPSRC (grants EP/V04902X/1, EP/R010161/1 and EP/V009680/1). E. Mart\'{\i}nez-Pa\~neda additionally acknowledges financial support from UKRI’s Future Leaders Fellowship programme [grant MR/V024124/1], and from the UKRI Horizon Europe Guarantee programme (ERC Starting Grant \textit{ResistHfracture}, EP/Y037219/1).

\section*{Data availability}

Data will be made available on request.

%%\newpage
\appendix

\section{Grain boundary characterization}
\label{appendix}
\setcounter{table}{0}

The $\Sigma$ values were identified according to Brandon’s criterion \begin{equation}
\Delta\theta = \frac{15^\circ}{\sqrt{\Sigma}}
\label{eq:tolerance}
\end{equation}

Where the misorientation angle \( \theta \) of the fractured GBs was calculated based on rotation matrices of neighbor grains \( R_1 \), \( R_2 \) and their relative rotation matrix \(R_{rel}\) by \begin{equation}
R_{rel} = R_2 \cdot R_1^T
\label{eq:relative_rotation}
\end{equation} \begin{equation}
\cos\left(\frac{\theta}{2}\right) = \frac{\text{Tr}(R_{rel}) - 1}{2}
\label{eq:misorientation_angle}
\end{equation} employing the MTEX MATLAB add-on package (version 5.10.0).

The $\Sigma$ value was determined based on misorientation angles reported in  \cite{Zhao1988}. Only $\Sigma$-3, $\Sigma$-5, $\Sigma$-7, $\Sigma$-9, $\Sigma$-23, $\Sigma$-29, $\Sigma$-31 and $\Sigma$-35 GBs exist in the baseline microstructure as CSL boundaries (Table \ref{table:CSL-misorientation-tolerance}). Therefore, these $\Sigma$ values were possible for fractured GBs. Additionally, GBs that failed to meet such criteria, i.e., random GBs, were categorized into general GBs.

\begin{table}[h!]
\centering
\caption{$\Sigma$ values of the identfied CSL grain boundaries with their \(\theta\) and \(\Delta\theta\) for cubic lattices (m$\overline{3}$m point group).}
\begin{tabular}{c c c  c c c}
\toprule
\textbf{$\Sigma$ Value} & \textbf{\(\theta\) ($^\circ$)} & \textbf{\(\Delta\theta\) ($^\circ$)} & \textbf{$\Sigma$ Value} & \textbf{\(\theta\) ($^\circ$)} & \textbf{\(\Delta\theta\) ($^\circ$)} \\
\midrule
3  & 60.00 & 8.66 & 23 & 40.45 & 3.13 \\
5  & 36.86 & 6.71 & 29 & 46.40 & 2.79 \\
7  & 38.21 & 5.67 & 31 & 52.20 & 2.6\\
9  & 38.94  & 5.00 & 35 & 34.05  & 2.54 \\
\bottomrule
\end{tabular}
\label{table:CSL-misorientation-tolerance}
\end{table}





% \end{thebibliography}

\small
\newpage
\begin{thebibliography}{10}
\expandafter\ifx\csname url\endcsname\relax
  \def\url#1{\texttt{#1}}\fi
\expandafter\ifx\csname urlprefix\endcsname\relax\def\urlprefix{URL }\fi
\expandafter\ifx\csname href\endcsname\relax
  \def\href#1#2{#2} \def\path#1{#1}\fi

\bibitem{Li2020}
X.~Li, X.~Ma, J.~Zhang, E.~Akiyama, Y.~Wang, X.~Song, Review of hydrogen
  embrittlement in metals: Hydrogen diffusion, hydrogen characterization,
  hydrogen embrittlement mechanism and prevention, Acta Metallurgica Sinica
  (English Letters) 33 (2020) 759--773.
\newblock \href {https://doi.org/10.1007/s40195-020-01039-7}
  {\path{doi:10.1007/s40195-020-01039-7}}.

\bibitem{Donovan1976}
J.~A. Donovan, Accelerated evolution of hydrogen from metals during plastic
  deformation, Metallurgical Transactions A 7 (1976) 1677--1683.
\newblock \href {https://doi.org/10.1007/BF02817885}
  {\path{doi:10.1007/BF02817885}}.

\bibitem{Martin2012}
M.~L. Martin, B.~P. Somerday, R.~O. Ritchie, P.~Sofronis, I.~M. Robertson,
  Hydrogen-induced intergranular failure in nickel revisited, Acta Materialia
  60 (2012) 2739--2745.
\newblock \href {https://doi.org/10.1016/J.ACTAMAT.2012.01.040}
  {\path{doi:10.1016/J.ACTAMAT.2012.01.040}}.

\bibitem{Ferreira1998}
P.~J. Ferreira, I.~M. Robertson, H.~K. Birnbaum, Hydrogen effects on the
  interaction between dislocations, Acta Materialia 46 (1998) 1749--1757.
\newblock \href {https://doi.org/10.1016/S1359-6454(97)00349-2}
  {\path{doi:10.1016/S1359-6454(97)00349-2}}.

\bibitem{Lu2020}
X.~Lu, Y.~Ma, D.~Wang, On the hydrogen embrittlement behavior of nickel-based
  alloys: Alloys 718 and 725, Materials Science and Engineering: A 792 (2020)
  139785.
\newblock \href {https://doi.org/10.1016/J.MSEA.2020.139785}
  {\path{doi:10.1016/J.MSEA.2020.139785}}.

\bibitem{Kumar2017}
B.~S. Kumar, V.~Kain, M.~Singh, B.~Vishwanadh, Influence of hydrogen on
  mechanical properties and fracture of tempered 13wt$\%$ \uppercase{C}r
  martensitic stainless steel, Materials Science and Engineering: A 700 (2017)
  140--151.
\newblock \href {https://doi.org/10.1016/J.MSEA.2017.05.086}
  {\path{doi:10.1016/J.MSEA.2017.05.086}}.

\bibitem{Tabata1984}
T.~Tabata, H.~K. Birnbaum, Direct observations of hydrogen enhanced crack
  propagation in iron, Scripta Metallurgica 18 (1984) 231--236.
\newblock \href {https://doi.org/10.1016/0036-9748(84)90513-1}
  {\path{doi:10.1016/0036-9748(84)90513-1}}.

\bibitem{Djukic2019}
M.~B. Djukic, G.~M. Bakic, V.~S. Zeravcic, A.~Sedmak, B.~Rajicic, The
  synergistic action and interplay of hydrogen embrittlement mechanisms in
  steels and iron: Localized plasticity and decohesion, Engineering Fracture
  Mechanics 216 (2019) 106528.
\newblock \href {https://doi.org/10.1016/J.ENGFRACMECH.2019.106528}
  {\path{doi:10.1016/J.ENGFRACMECH.2019.106528}}.

\bibitem{Singh2024MatSciA}
V.~Singh, A.~Raj, D.~K. Mahajan, Investigation into hydrogen assisted fracture
  in nickel oligocrystals, Materials Science and Engineering: A 895 (2024)
  146257.
\newblock \href {https://doi.org/10.1016/J.MSEA.2024.146257}
  {\path{doi:10.1016/J.MSEA.2024.146257}}.

\bibitem{Soundararajan2023}
C.~K. Soundararajan, A.~Myhre, A.~Sendrowicz, X.~Lu, A.~Vinogradov,
  Hydrogen-induced degradation behavior of nickel alloy studied using acoustic
  emission technique, Materials Science and Engineering: A 865 (2023) 144635.
\newblock \href {https://doi.org/10.1016/J.MSEA.2023.144635}
  {\path{doi:10.1016/J.MSEA.2023.144635}}.

\bibitem{Liu2024MatSciA}
Q.~Liu, S.~Manzoor, Y.~Yan, M.~Tariq, A.~Saul, H.~Farhat, A.~Barnoush,
  Influence of hydrogen uptake on additive manufacturing and conventional
  austenitic stainless steels 316\uppercase{L}, Materials Science and
  Engineering: A 914 (2024) 147170.
\newblock \href {https://doi.org/10.1016/J.MSEA.2024.147170}
  {\path{doi:10.1016/J.MSEA.2024.147170}}.

\bibitem{Robertson1984}
I.~M. Robertson, T.~Tabata, W.~Wei, F.~Heubaum, H.~K. Birnbaum, Hydrogen
  embrittlement and grain boundary fracture, Scripta Metallurgica 18 (1984)
  841--846.
\newblock \href {https://doi.org/10.1016/0036-9748(84)90407-1}
  {\path{doi:10.1016/0036-9748(84)90407-1}}.

\bibitem{Harris2018}
Z.~D. Harris, S.~K. Lawrence, D.~L. Medlin, G.~Guetard, J.~T. Burns, B.~P.
  Somerday, Elucidating the contribution of mobile hydrogen-deformation
  interactions to hydrogen-induced intergranular cracking in polycrystalline
  nickel, Acta Materialia 158 (2018) 180--192.
\newblock \href {https://doi.org/10.1016/j.actamat.2018.07.043}
  {\path{doi:10.1016/j.actamat.2018.07.043}}.

\bibitem{Bechtle2009}
S.~Bechtle, M.~Kumar, B.~P. Somerday, M.~E. Launey, R.~O. Ritchie,
  Grain-boundary engineering markedly reduces susceptibility to intergranular
  hydrogen embrittlement in metallic materials, Acta Materialia 57 (2009)
  4148--4157.
\newblock \href {https://doi.org/10.1016/j.actamat.2009.05.012}
  {\path{doi:10.1016/j.actamat.2009.05.012}}.

\bibitem{Lee1986}
S.-M. Lee, J.-Y. Lee, The trapping and transport phenomena of hydrogen in
  nickel, Metallurgical Transactions A 17 (1986) 181--187.
\newblock \href {https://doi.org/10.1007/BF02643893}
  {\path{doi:10.1007/BF02643893}}.

\bibitem{Chen2024}
Y.~S. Chen, C.~Huang, P.~Y. Liu, H.~W. Yen, R.~Niu, P.~Burr, K.~L. Moore,
  E.~Martínez-Pañeda, A.~Atrens, J.~M. Cairney, Hydrogen trapping and
  embrittlement in metals – a review, International Journal of Hydrogen
  Energy (4 2024).
\newblock \href {https://doi.org/10.1016/J.IJHYDENE.2024.04.076}
  {\path{doi:10.1016/J.IJHYDENE.2024.04.076}}.

\bibitem{Oudriss2012}
A.~Oudriss, J.~Creus, J.~Bouhattate, E.~Conforto, C.~Berziou, C.~Savall,
  X.~Feaugas, Grain size and grain-boundary effects on diffusion and trapping
  of hydrogen in pure nickel, Acta Materialia 60 (2012) 6814--6828.
\newblock \href {https://doi.org/10.1016/J.ACTAMAT.2012.09.004}
  {\path{doi:10.1016/J.ACTAMAT.2012.09.004}}.

\bibitem{Wada2023}
K.~Wada, C.~Shibata, H.~Enoki, T.~Iijima, J.~Yamabe, Hydrogen-induced
  intergranular cracking of pure nickel under various strain rates and
  temperatures in gaseous hydrogen environment, Materials Science and
  Engineering: A 873 (2023) 145040.
\newblock \href {https://doi.org/10.1016/j.msea.2023.145040}
  {\path{doi:10.1016/j.msea.2023.145040}}.

\bibitem{Wada2019}
K.~Wada, J.~Yamabe, H.~Matsunaga, Visualization of trapped hydrogen along grain
  boundaries and its quantitative contribution to hydrogen-induced
  intergranular fracture in pure nickel, Materialia 8 (2019) 100478.
\newblock \href {https://doi.org/10.1016/J.MTLA.2019.100478}
  {\path{doi:10.1016/J.MTLA.2019.100478}}.

\bibitem{Hachet2021}
G.~Hachet, A.~Oudriss, A.~Barnoush, T.~Hajilou, D.~Wang, A.~Metsue, X.~Feaugas,
  Antagonist softening and hardening effects of hydrogen investigated using
  nanoindentation on cyclically pre-strained nickel single crystal, Materials
  Science and Engineering: A 803 (2021) 140480.
\newblock \href {https://doi.org/10.1016/J.MSEA.2020.140480}
  {\path{doi:10.1016/J.MSEA.2020.140480}}.

\bibitem{Jothi2016}
S.~Jothi, S.~V. Merzlikin, T.~N. Croft, J.~Andersson, S.~G. Brown, An
  investigation of micro-mechanisms in hydrogen induced cracking in
  nickel-based superalloy 718, Journal of Alloys and Compounds 664 (2016)
  664--681.
\newblock \href {https://doi.org/10.1016/J.JALLCOM.2016.01.033}
  {\path{doi:10.1016/J.JALLCOM.2016.01.033}}.

\bibitem{Sangid2010}
M.~D. Sangid, H.~Sehitoglu, H.~J. Maier, T.~Niendorf, Grain boundary
  characterization and energetics of superalloys, Materials Science and
  Engineering: A 527 (2010) 7115--7125.
\newblock \href {https://doi.org/10.1016/J.MSEA.2010.07.062}
  {\path{doi:10.1016/J.MSEA.2010.07.062}}.

\bibitem{Harris2019MatSciA}
Z.~D. Harris, J.~T. Burns, The effect of isothermal heat treatment on hydrogen
  environment-assisted cracking susceptibility in monel \uppercase{K}-500,
  Materials Science and Engineering: A 764 (2019) 138249.
\newblock \href {https://doi.org/10.1016/J.MSEA.2019.138249}
  {\path{doi:10.1016/J.MSEA.2019.138249}}.

\bibitem{Brandon1966}
D.~G. Brandon, The structure of high-angle grain boundaries, Acta Metallurgica
  14 (1966) 1479--1484.
\newblock \href {https://doi.org/10.1016/0001-6160(66)90168-4}
  {\path{doi:10.1016/0001-6160(66)90168-4}}.

\bibitem{Tehranchi2017-II}
A.~Tehranchi, W.~A. Curtin, Atomistic study of hydrogen embrittlement of grain
  boundaries in nickel: \textnormal{II}. decohesion, Modelling and Simulation
  in Materials Science and Engineering 25 (2017) 075013.
\newblock \href {https://doi.org/10.1088/1361-651X/aa87a6}
  {\path{doi:10.1088/1361-651X/aa87a6}}.

\bibitem{Mai2021}
H.~L. Mai, X.~Y. Cui, D.~Scheiber, L.~Romaner, S.~P. Ringer, An understanding
  of hydrogen embrittlement in nickel grain boundaries from first principles,
  Materials \& Design 212 (2021) 110283.
\newblock \href {https://doi.org/10.1016/J.MATDES.2021.110283}
  {\path{doi:10.1016/J.MATDES.2021.110283}}.

\bibitem{Li2020DFT}
J.~Li, C.~Lu, L.~Pei, C.~Zhang, R.~Wang, Atomistic investigation of hydrogen
  induced decohesion of \uppercase{N}i grain boundaries, Mechanics of Materials
  150 (2020) 103586.
\newblock \href {https://doi.org/10.1016/J.MECHMAT.2020.103586}
  {\path{doi:10.1016/J.MECHMAT.2020.103586}}.

\bibitem{DiStefanoDavide2015}
D.~D. Stefano, M.~Mrovec, C.~Elsässer, First-principles investigation of
  hydrogen trapping and diffusion at grain boundaries in nickel, Acta
  Materialia 98 (2015) 306--312.
\newblock \href {https://doi.org/10.1016/j.actamat.2015.07.031}
  {\path{doi:10.1016/j.actamat.2015.07.031}}.

\bibitem{Alvaro2015}
A.~Alvaro, I.~T. Jensen, N.~Kheradmand, O.~M. Løvvik, V.~Olden, Hydrogen
  embrittlement in nickel, visited by first principles modeling, cohesive zone
  simulation and nanomechanical testing, International Journal of Hydrogen
  Energy 40 (2015) 16892--16900.
\newblock \href {https://doi.org/10.1016/j.ijhydene.2015.06.069}
  {\path{doi:10.1016/j.ijhydene.2015.06.069}}.

\bibitem{Hajilou2020}
T.~Hajilou, I.~Taji, F.~Christien, S.~He, D.~Scheiber, W.~Ecker, R.~Pippan,
  V.~I. Razumovskiy, A.~Barnoush, Hydrogen-enhanced intergranular failure of
  sulfur-doped nickel grain boundary: In situ electrochemical micro-cantilever
  bending vs. dft, Materials Science and Engineering: A 794 (2020) 139967.
\newblock \href {https://doi.org/10.1016/J.MSEA.2020.139967}
  {\path{doi:10.1016/J.MSEA.2020.139967}}.

\bibitem{Li2019}
J.~Li, C.~Lu, L.~Pei, C.~Zhang, R.~Wang, K.~Tieu, Influence of hydrogen
  environment on dislocation nucleation and fracture response of 〈110〉
  grain boundaries in nickel, Computational Materials Science 165 (2019)
  40--50.
\newblock \href {https://doi.org/10.1016/J.COMMATSCI.2019.04.027}
  {\path{doi:10.1016/J.COMMATSCI.2019.04.027}}.

\bibitem{Ding2022}
Y.~Ding, H.~Yu, M.~Lin, K.~Zhao, S.~Xiao, A.~Vinogradov, L.~Qiao, M.~Ortiz,
  J.~He, Z.~Zhang, Hydrogen-enhanced grain boundary vacancy stockpiling causes
  transgranular to intergranular fracture transition, Acta Materialia 239
  (2022) 118279.
\newblock \href {https://doi.org/10.1016/j.actamat.2022.118279}
  {\path{doi:10.1016/j.actamat.2022.118279}}.

\bibitem{Takahashi2016}
Y.~Takahashi, H.~Kondo, R.~Asano, S.~Arai, K.~Higuchi, Y.~Yamamoto, S.~Muto,
  N.~Tanaka, Direct evaluation of grain boundary hydrogen embrittlement: A
  micro-mechanical approach, Materials Science and Engineering: A 661 (2016)
  211--216.
\newblock \href {https://doi.org/10.1016/J.MSEA.2016.03.035}
  {\path{doi:10.1016/J.MSEA.2016.03.035}}.

\bibitem{Seita2015}
M.~Seita, J.~P. Hanson, S.~Gradečak, M.~J. Demkowicz, The dual role of
  coherent twin boundaries in hydrogen embrittlement, Nature Communications 6
  (2015) 6164.
\newblock \href {https://doi.org/10.1038/ncomms7164}
  {\path{doi:10.1038/ncomms7164}}.

\bibitem{Zhang2020}
Z.~Zhang, Z.~Yang, S.~Lu, A.~Harte, R.~Morana, M.~Preuss, Strain localisation
  and failure at twin-boundary complexions in nickel-based superalloys, Nature
  Communications 11 (2020) 4890.
\newblock \href {https://doi.org/10.1038/s41467-020-18641-z}
  {\path{doi:10.1038/s41467-020-18641-z}}.

\bibitem{Liu2024}
M.~Liu, L.~Jiang, M.~J. Demkowicz, Role of slip in hydrogen-assisted crack
  initiation in \uppercase{N}i-based alloy 725, Science Advances 10~(29) (2024)
  2118.
\newblock \href {https://doi.org/10.1126/sciadv.ado2118}
  {\path{doi:10.1126/sciadv.ado2118}}.

\bibitem{Livia2024}
L.~Cupertino-Malheiros, M.~Duportal, T.~Hageman, A.~Zafra,
  E.~Martínez-Pañeda, Hydrogen uptake kinetics of cathodic polarized metals
  in aqueous electrolytes, Corrosion Science 231 (2024) 111959.
\newblock \href {https://doi.org/10.1016/J.CORSCI.2024.111959}
  {\path{doi:10.1016/J.CORSCI.2024.111959}}.

\bibitem{Lu2019}
X.~Lu, D.~Wang, D.~Wan, Z.~B. Zhang, N.~Kheradmand, A.~Barnoush, Effect of
  electrochemical charging on the hydrogen embrittlement susceptibility of
  alloy 718, Acta Materialia 179 (2019) 36--48.
\newblock \href {https://doi.org/10.1016/J.ACTAMAT.2019.08.020}
  {\path{doi:10.1016/J.ACTAMAT.2019.08.020}}.

\bibitem{Pérez2011}
D.~P. Escobar, C.~Miñambres, L.~Duprez, K.~Verbeken, M.~Verhaege, Internal and
  surface damage of multiphase steels and pure iron after electrochemical
  hydrogen charging, Corrosion Science 53 (2011) 3166--3176.
\newblock \href {https://doi.org/10.1016/J.CORSCI.2011.05.060}
  {\path{doi:10.1016/J.CORSCI.2011.05.060}}.

\bibitem{Hanson2018}
J.~P. Hanson, A.~Bagri, J.~Lind, P.~Kenesei, R.~M. Suter, S.~Gradečak, M.~J.
  Demkowicz, Crystallographic character of grain boundaries resistant to
  hydrogen-assisted fracture in \uppercase{N}i-base alloy 725, Nature
  Communications 9 (2018) 3386.
\newblock \href {https://doi.org/10.1038/s41467-018-05549-y}
  {\path{doi:10.1038/s41467-018-05549-y}}.

\bibitem{Kwon2018}
Y.~J. Kwon, S.~P. Jung, B.~J. Lee, C.~S. Lee, Grain boundary engineering
  approach to improve hydrogen embrittlement resistance in
  \uppercase{F}e-\uppercase{M}n-\uppercase{C} \uppercase{TWIP} steel,
  International Journal of Hydrogen Energy 43 (2018) 10129--10140.
\newblock \href {https://doi.org/10.1016/J.IJHYDENE.2018.04.048}
  {\path{doi:10.1016/J.IJHYDENE.2018.04.048}}.

\bibitem{Ma2021}
Z.~Ma, X.~Xiong, Y.~Su, Study on hydrogen segregation at individual grain
  boundaries in pure nickel by scanning kelvin probe force microscopy,
  Materials Letters 303 (2021) 130528.
\newblock \href {https://doi.org/10.1016/J.MATLET.2021.130528}
  {\path{doi:10.1016/J.MATLET.2021.130528}}.

\bibitem{Lassila1988}
D.~H. Lassila, H.~K. Birnbaum, The effect of diffusive segregation on the
  fracture of hydrogen charged nickel, Acta Metallurgica 36 (1988) 2821--2825.
\newblock \href {https://doi.org/10.1016/0001-6160(88)90128-9}
  {\path{doi:10.1016/0001-6160(88)90128-9}}.

\bibitem{Fernndez-Sousa2020}
R.~Fernández-Sousa, C.~Betegón, E.~Martínez-Pañeda, Analysis of the
  influence of microstructural traps on hydrogen assisted fatigue, Acta
  Materialia 199 (2020) 253--263.
\newblock \href {https://doi.org/10.1016/J.ACTAMAT.2020.08.030}
  {\path{doi:10.1016/J.ACTAMAT.2020.08.030}}.

\bibitem{Tehranchi2020}
A.~Tehranchi, X.~Zhou, W.~A. Curtin, A decohesion pathway for hydrogen
  embrittlement in nickel: Mechanism and quantitative prediction, Acta
  Materialia 185 (2020) 98--109.
\newblock \href {https://doi.org/10.1016/J.ACTAMAT.2019.11.062}
  {\path{doi:10.1016/J.ACTAMAT.2019.11.062}}.

\bibitem{Chen2013}
S.~Chen, M.~Zhao, L.~Rong, Hydrogen-induced cracking behavior of twin boundary
  in $\gamma'$ phase strengthened \uppercase{F}e–\uppercase{N}i based
  austenitic alloys, Materials Science and Engineering: A 561 (2013) 7--12.
\newblock \href {https://doi.org/10.1016/J.MSEA.2012.10.069}
  {\path{doi:10.1016/J.MSEA.2012.10.069}}.

\bibitem{Zhao1988}
J.~Zhao, B.~L. Adams, Definition of an asymmetric domain for intercrystalline
  misorientation in cubic materials in the space of euler angles, Acta
  Crystallographica Section A 44 (1988) 326--336.
\newblock \href {https://doi.org/10.1107/S010876738701256X}
  {\path{doi:10.1107/S010876738701256X}}.

\end{thebibliography}
\end{document}